\documentclass[aps,prx,twocolumn,footinbib]{revtex4-2}
\usepackage{graphicx}
\usepackage{indentfirst}
\usepackage{physics}
\usepackage{braket}
\usepackage{float}
\usepackage{amsmath}
\usepackage{epstopdf}
\usepackage{footnote} 
\usepackage{CJK}
\usepackage{esint}
\usepackage{color}
\usepackage[T1]{fontenc}
\usepackage{subfigure}
\usepackage{amsfonts}
\usepackage{footmisc}
\usepackage{scrextend}
\usepackage{multirow}
\usepackage[hyperfootnotes=false]{hyperref}
\usepackage[acronym]{glossaries}

\usepackage[english]{babel}
\usepackage{url}
\usepackage{bm}
\usepackage{hyperref}
\definecolor{darkblue}{rgb}{0,0,0.5}
\hypersetup{
colorlinks=true,
linkcolor=black,
filecolor=blue,
citecolor=darkblue,  
urlcolor=black,
}

\newcommand{\calL}{{\cal L}}

\newcommand{\1}{^{(1)}}

\def\be{\begin{equation}}
\def\ee{\end{equation}}
\def\ba{\begin{eqnarray}}
\def\ea{\end{eqnarray}}
\usepackage{bm}

\newcommand{\QZ}[1]{{{\textcolor{black}{#1}}}}

\begin{document}

%title to be done 
\title{Quantum ranging with Gaussian entanglement}

\author{Quntao Zhuang}
\email{zhuangquntao@email.arizona.edu}

\address{
Department of Electrical and Computer Engineering \& 
James C. Wyant College of Optical Sciences,
\\
 University of Arizona, Tucson, AZ 85721, USA
}

\begin{abstract}
It is well known that entanglement can benefit quantum information processing tasks. Quantum illumination, when first proposed, is surprising as entanglement's benefit survives entanglement-breaking noise. Since then, many efforts have been devoted to study quantum sensing in noisy scenarios. The applicability of such schemes, however, is limited to a binary quantum hypothesis testing scenario. In terms of target detection, such schemes interrogate a single polarization-azimuth-elevation-range-Doppler resolution bin at a time, limiting the impact to radar detection. We resolve this \QZ{binary-hypothesis limitation} by proposing a quantum ranging protocol enhanced by entanglement. By formulating a ranging task as a multiary hypothesis testing problem, we show that entanglement enables a 6-dB advantage in the error exponent against the optimal classical scheme. Moreover, the proposed ranging protocol can also be \QZ{utilized} to implement a pulse-position modulated entanglement-assisted communication protocol. Our ranging protocol reveals entanglement's potential in general quantum hypothesis testing tasks and paves the way towards a \QZ{quantum-ranging} radar with a provable quantum advantage. 

\end{abstract} 
\maketitle

%\section{Introduction}

Entanglement is one of the most intriguing phenomena promised by quantum physics. As the ``spooky
action at a distance'' unveils itself with the development of quantum physics, entanglement also turns out to be beneficial to various applications in communication~\cite{gisin2002,xu2020,pirandola2020advances}, computation~\cite{Preskill2018quantumcomputingin} and sensing~\cite{giovannetti2011advances,degen2017quantum,braun2018rmp,pirandola2018advances,sidhu2020geometric}. In computation, entangling multiple qubits in a well-controlled manner enables the efficient computation of difficult problems~\cite{Shor_1997}. In communication, entanglement enables a higher information transmission rate~\cite{bennett1992,bennett2002entanglement} and provides unconditional security~\cite{Bennett20147,Ekert_1991}. In sensing, entanglement enables the Heisenberg scaling~\cite{zwierz2010general} in measuring an identical parameter among sensors~\cite{giovannetti2006} or even a global property of parameters distributed across different sensors~\cite{ge2017distributed,proctor2017multi,zhuang2018distributed,eldredge2018optimal,zhang2020distributed}. 

Entanglement is fragile---noise and loss can easily destroy it, yet surprisingly its operational advantages can survive. For example, the rate of entanglement-assisted (EA) communication can be much larger than the un-assisted classical capacity, even for an entanglement-breaking channel that destroys the entanglement at the receiver side, as predicted by the theory works~\cite{bennett2002entanglement,shi2020practical} and recently demonstrated in an experiment~\cite{hao2020}. In quantum illumination (QI)~\cite{Lloyd2008,tan2008quantum}, the target's presence can be probed with a 6dB advantage in the error exponent, despite the original entanglement being entirely destroyed at the receiver side.

Many efforts have been devoted to make QI's theoretical advantage practically relevant. Sub-optimal receiver designs~\cite{Guha2009} that enable experimental demonstrations~\cite{zhang2013,zhang2015,Lopaeva_2013} and a structured optimal receiver design to saturate the quantum advantage~\cite{zhuang2017} have been proposed. To adapt to a radar detection scenario, extensions to Neyman-Pearson decision strategy~\cite{zhuang2017NP} and target fading scenarios~\cite{zhuang2017fading} have been achieved. As the large noise background required by QI's advantage exists only in microwave, demonstration in the microwave domain is also an overall goal~\cite{barzanjeh2015microwave,barzanjeh2020microwave,chang2019quantum}. 
However, as pointed out in recent reviews~\cite{pirandola2018advances,shapiro2020quantum}, a major hurdle that prevents QI being eventually practically advantageous is its limitation to be only able to interrogate a single polarization-azimuth-elevation-range-Doppler resolution bin at a time. Despite recent theoretical advances in multiary channel discrimination~\cite{zhuang2020entanglement,zhuang2020ultimate} that bring hope to solve the problem, energetic considerations seem to show that no entanglement advantage can be obtained~\cite{karsa2020energetic} from that perspective.

%More importantly, as in the classical case, only a single pulse concentrating all probe energy is necessary for the task, energetic considerations show that no entanglement advantage can be obtained~\cite{karsa2020energetic}. 

In this letter, we resolve the \QZ{limitation} by proposing a quantum ranging protocol enhanced by Gaussian entanglement~\cite{Weedbrook2012}. First, to go beyond previous studies~\cite{zhuang2020entanglement}, we develop a precise model for the ranging task, where one sends out a signal pulse and continuously measure at the receiver side to determine the reflection of a target at line-of-sight. As any ranging task has a finite precision requirement, we then formulate ranging as a multiary hypothesis testing problem, where each hypothesis corresponds to target being in one of the $m\ge 2$ slices of discretized range. 
We show that by storing an idler entangled with the signal pulse, the target range can be determined with a
6dB advantage in the error exponent. Our results on quantum ranging also directly apply to a pulse-position modulated EA classical communication protocol, offering a rate much higher than the classical capacity in the low-photon number region. We design a practical receiver in the $m=2$ case that enables entanglement advantage and provide intuition for the optimal receiver in the general case. 

%. Choosing the two-mode squeezed vacuum (TMSV) state as the entangled source, we evalute the quantum Chernoff bound (QCB) of the multi-hypothesis testing problem~\cite{li2016discriminating,nussbaum2011asymptotic,audenaert2007discriminating} to prove a 

\begin{figure}[t]
    \centering
    \includegraphics[width=0.4\textwidth]{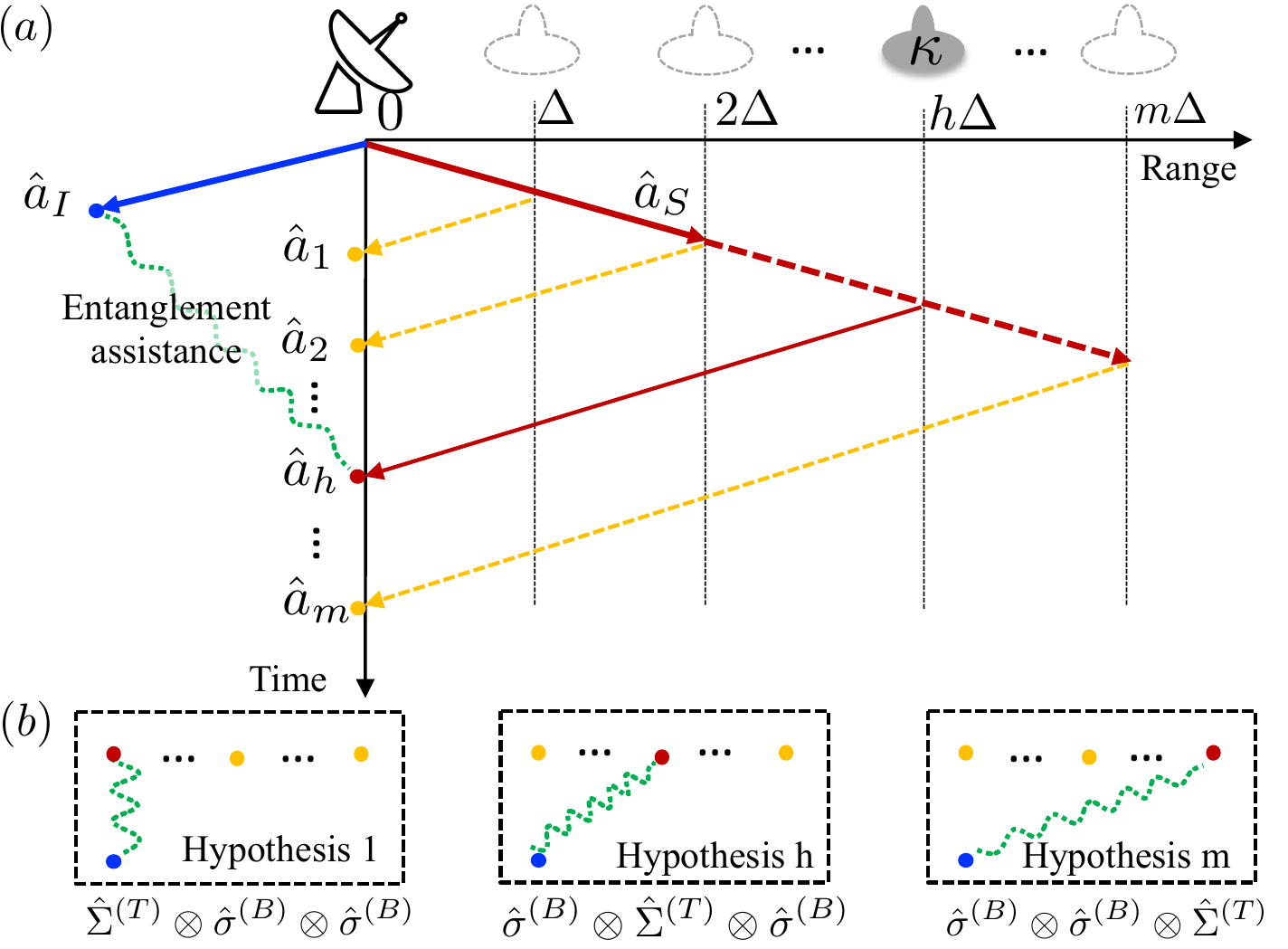}
    \caption{
    Schematic of the entanglement-assisted ranging protocol. In (a), the signal mode $\hat{a}_S$ (blue) and the idler mode $\hat{a}_I$ (red) are initially entangled in a TMSV state. The signal is sent out to probe the range of a target with reflectivity $\kappa$. When the target is at distance $h\Delta$, the mode $\hat{a}_h$ highlighted in red collected at time $t_h=2h\Delta/c$ contains the reflection from the target embedded in noise, while the rest of the collected modes (orange) contain entirely noise. Subplot (b) shows the $m$ possible states in the hypothesis testing problem at the receiver side. In each case, the idler (blue) is correlated with the reflected mode (red).
    \label{fig:schematic}
    }
\end{figure}

{\em Model of ranging.---}
We consider the task of determining the distance between an observer and a target along the line-of-sight. Suppose the observer has a finite precision requirement $\Delta$, then we can divide the line-of-sight into $m\ge 2$ length-$\Delta$ slices, and model the problem of ranging as a hypothesis testing task between $m$ hypotheses (see Fig.~\ref{fig:schematic}). In hypothesis $h$, the target is present in the slice centered at the position $h\Delta$ from the origin. 

To determine the range, one can send out a pulse, described by the mode annihilation operator $\hat{a}_S$, and wait for the reflected return from the target. The mean photon number of the mode $\braket{\hat{a}_S^\dagger \hat{a}_S}=N_S$ is constrained by the source brightness or to avoid revealing the attempt of detection. To determine the time of arrival of the returned pulse, one needs to continuously collect light at the receiver side, obtaining the modes $\{\hat{a}_{\ell}\}_{\ell=1}^m$, each arriving at time $t_\ell=2\ell \Delta/c$. 
In hypothesis $h$, the target is $h\Delta$ away from the observer, and the reflected mode $\hat{a}_{h}$ arrives at the observer after time $t_h=2h\Delta/c$. We can model the reflection by a bosonic thermal-loss channel $\calL_{\kappa,N_B}$ described by the beamsplitter transform
\be 
\hat{a}_{h}=\sqrt{\kappa} \hat{a}_S+\sqrt{1-\kappa}\hat{e}_h,
\label{eq:ah}
\ee 
where $\kappa$ is the target reflectivity and the noise mode $\hat{e}_h$ is in a thermal state with $N_B/(1-\kappa)$ photons on average.
When the returned signal does not arrive at time $t_\ell$, the noise mode being collected $\hat{a}_{\ell \neq h}=\hat{e}_\ell$ is in a thermal state with mean photon number $N_B$.

%, which is in a zero-mean Gaussian state described by the covariance matrix $(2N_B+1){\bm I}_2$ with ${\mathbf I}_2$ being the identity matrix~\cite{Weedbrook2012}. 

Now the task of ranging has been reduced to the determination of the returned signal mode $\hat{a}_h$ among the entire set of collected modes $\{\hat{a}_{\ell}\}_{\ell=1}^m$. 
In a classical scheme, the input state of $\hat{a}_S$ is assumed to have a positive P-function, as widely considered in the literature~\cite{tan2008quantum,pirandola2011quantum,zhuang2020entanglement}. 
In an entangled scheme, besides sending over the energy-constrained signal mode $\hat{a}_S$, one can also keep a locally-stored idler $\hat{a}_I$ entangled with the signal as depicted in Fig.~\ref{fig:schematic}. Similar to the case of QI~\cite{nair2020fundamental,bradshaw2020optimal}, we consider the signal-idler pair in the two-mode squeezed vacuum (TMSV) state~(see Appendix A), which we expect to be optimal. As depicted in Fig.~\ref{fig:schematic}(b), the stored idler mode $\hat{a}_I$ will still be correlated with the signal mode $\hat{a}_h$ returned from the thermal-loss channel $\calL_{\kappa,N_B}$ in hypothesis $h$, although the initial entanglement might be destroyed. The joint state of $\hat{a}_S$ and $\hat{a}_h$ has the covariance matrix
\begin{align}
&
{\mathbf{V}}_{SI}^\prime=
\left(
\begin{array}{cccc}
(2N_B+1) {\mathbf I}_2&2\sqrt{\kappa}C_p{\mathbf Z}_2\\
2\sqrt{\kappa}C_p{\mathbf Z}_2&(2N_S+1){\mathbf I}_2
\end{array} 
\right),
\label{noisy_cov}
&
\end{align}
where $C_p=\sqrt{N_S\left(N_S+1\right)}$, ${\mathbf I}_2$ and ${\mathbf Z}_2$ are the Pauli matrices. \QZ{Here we have chosen the unit such that the vacuum noise is unity. As we have $\kappa\ll1$, we have omitted the brightness signature in the signal; Note that the results are similar even if we include this difference.}

From the potential correlation depicted in Fig.~\ref{fig:schematic}(b), it is clear that ranging does not belong to the problem of quantum channel position finding (CPF) defined in Ref.~\cite{zhuang2020entanglement}: in ranging, it is unclear which pair of signal-idler is potentially correlated, while in CPF the pairing between potential correlated signals and idlers are clear.

{\em Hypothesis testing analyses.---}
The performance of the above hypothesis testing task is quantified by the error probability. To obtain the best performance, one can optimize the input state, under the total photon number constraint $N_S$, and the corresponding measurement. One can also utilize multiple degrees of freedom and send over $M$ modes $\hat{\bm a}_S\equiv \{\hat{a}_S^{(n)}\}_{n=1}^M$ in each pulse, therefore each portion of collected light also contains multiple modes $\hat{\bm a}_\ell \equiv\{\hat{a}_{\ell}^{(n)}\}_{n=1}^M$ for each time slice $ t_\ell$.

In the classical strategy, conditioned on the target range being $h\Delta$, the output state can be written as
\be 
\hat{\rho}_h^C=\left(\otimes_{\ell\neq h} \hat{\sigma}^{(B)}_{\hat{\bm a}_{\ell}}\right)\otimes \hat{\sigma}^{(T)}_{\hat{\bm a}_{h}},
\label{rho_C}
\ee 
where the background state $\hat{\sigma}^{(B)}$ consists of a product of $M$ thermal states, each with mean photon number $N_B$, and the target state $\hat{\sigma}^{(T)}$ is the $M$-mode returned signal embedded in the same thermal background, produced by the thermal-loss channel $\calL_{\kappa,N_B}$ in Eq.~\eqref{eq:ah}. 
\begin{figure*}
    \centering
    \includegraphics[width=0.7\textwidth]{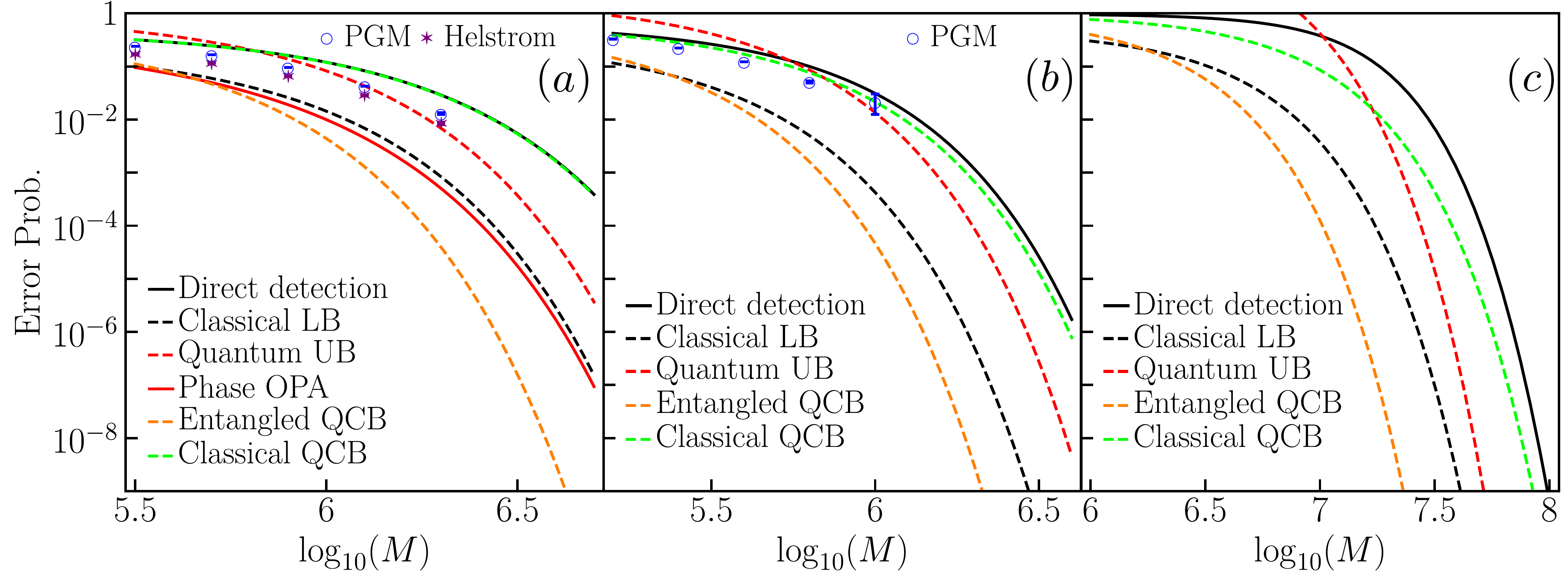}
    \caption{
    Error probability performance versus the number of modes $M$ of the quantum ranging protocol in comparison with the classical schemes. Signal brightness $N_S=0.001$ and target reflectivity \QZ{$\kappa=0.01$}. The number of range slices $m$ and the environmental noise $N_B$ are chosen as (a) $m=2, N_B=3$, (b) $m=3, N_B=1$ and (c) $m=50, N_B=20$. For the entangled strategy, we evaluate the asymptotically-tight quantum Chernoff bound (QCB) $P_{E,H}$ (orange dashed), and an exact upper bound $P_{E,UB}$ (red dashed); For the classical strategy, we evaluate the QCB $P_{C,H}$ (green dashed), an exact lower bound $P_{C,LB}$ (black dashed) and the coherent-state direction detection performance $P_{C,DD}$ (black solid). In (a), we also present the OPA-based receiver performance (red solid) in an entangled strategy and the numerical results of the classical Helstrom limit (purple star). In (a)(b), the performance of the pretty-good-measurement (PGM) for coherent-state inputs is also evaluated numerically in comparison to the classical QCB.
    \label{fig:m}
    }
\end{figure*}
In the entangled scheme, each signal mode $\hat{a}_S^{(n)}$ has an idler $\hat{a}_I^{(n)}$ stored locally, and the overall return-idler state is
\be 
\hat{\rho}_h^E=\left(\otimes_{\ell\neq h} \hat{\sigma}^{(B)}_{\hat{\bm a}_{\ell}}\right)\otimes \hat{\Sigma}^{(T)}_{\hat{\bm a}_{h}\hat{\bm a}_I},
\label{rho_E}
\ee 
where the correlated output state $\hat{\Sigma}^{(T)}$ has $M$ pairs of signal-idler, each in the state described by the covariance matrix ${\mathbf{V}}_{SI}^\prime$ in Eq.~\eqref{noisy_cov}.

Given the positive operator-valued measure (POVM) elements $\{\hat{\Pi}_\ell^{C/E}\}_{\ell=1}^m$ describing the measurement in the classical (C) or entangled (E) scheme, \QZ{with each element $\hat{\Pi}_\ell^{C/E}$ representing the decision that target range is $\ell \Delta$}, the error probability
$
P_{C/E}=1-\sum_{\ell=1}^m  p_\ell \tr\left[\hat{\Pi}^{C/E}_\ell\hat{\rho}_\QZ{\ell}^{C/E}\right],
$ 
where the priors $p_\ell=1/m$ can be chosen uniform without loss of generality.

{\em Performance of classical schemes.---} Utilizing the convexity of the Helstrom limit and the quantum Chernoff bound (QCB)~\cite{li2016discriminating,nussbaum2011asymptotic,audenaert2007discriminating,Pirandola2008}, we can derive an asymptotically tight expression of the error probability limit of any classical strategy utilizing inputs with a positive P-function~(see Appendix B)
\begin{align}
P_{C,H}&\sim  \frac{m-1}{m}\exp\left[-\frac{2M\kappa N_S}{1+2N_B+2\sqrt{N_B\left(1+N_B\right)}}\right]
\nonumber
\\
&\simeq 
\frac{m-1}{m}\exp\left[-\frac{M\kappa N_S}{2N_B}\right], \mbox{when $N_B\gg1$},
\label{P_C_H_QCB}
\end{align} 
which is \QZ{tight in the error exponent. Here the constant $(m-1)/m$ is chosen to match the low signal-to-noise ratio limit with random guess}. The limit is achieved by any coherent state input under the proper energy constraint.
Furthermore, despite ranging being different from CPF~\cite{zhuang2020entanglement}, as in the classical strategy no idlers are present, the ultimate lower bound of the error probability of classical CPF (Eq.~(10) in Ref.~\cite{zhuang2020entanglement}) also applies to classical ranging
$
P_{C,LB} = {(m-1)}/{2m}\times\exp\left[-{2MN_S\kappa}/{(1+2N_B)} \right],
$
however giving an error exponent twice larger than that of Eq.~\eqref{P_C_H_QCB}.

We can compare the asymptotic limit in Eq.~\eqref{P_C_H_QCB} with the error-probability performance of the single-mode coherent-state direct detection (DD) strategy~\cite{Helstrom_1976,cariolaro2010theory}
\begin{align}
P_{C,DD}=&\frac{1}{m}\sum_{k=2}^m (-1)^k C_m^k \cross \nonumber \\ 
& \exp\left[-\frac{(1-v)(1-v^{k-1})\kappa MN_S}{1-v^k}\right],
\label{P_CI_DD}
\end{align}
where $v=N_B/(N_B+1)$ and $C_m^k$ is the binomial coefficient (number of combinations of $k$ items out of $m$). In the high-noise $N_B\gg1$ and large number of modes $M\gg1$ limit,
$
P_{C,DD}\sim\exp\left(-M\kappa N_S/2N_B\right)
$.
We see that coherent-state DD is the asymptotic optimal classical strategy \QZ{in terms of the error exponent}.

We evaluate $P_{C,H}$ (green dashed), $P_{C,LB}$ (black dashed) and the exact version of $P_{C,DD}$ (black solid) in Fig.~\ref{fig:m} for various parameters. Indeed, we see that $P_{C,DD}$ collapses with $P_{C,H}$ for $m=2$ (subplot (a)) and asymptotically agrees with $P_{C,H}$ even for $m>2$ (subplots (b)(c)). We also numerically evaluate the Helstrom limit in the $m=2$ case, and find that $P_{C,H}$ indeed provides the correct scaling. For $m>2$, numerical evaluation of the Helstrom limit is challenging, we compare with the performance of the pretty-good measurement (PGM)~\cite{PGM1,PGM2,PGM3,zhuang2020entanglement}, which agrees well with the Helstrom limit in the $m=2$ case in Fig.~\ref{fig:m}(a). For the $m=3$ case, a good agreement between the PGM performance and $P_{C,H}$ can be seen. Therefore, we conclude that $P_{C,H}$ and $P_{C,DD}$ well characterize the classical performance limit.

{\em Entanglement advantage.---}In the EA ranging protocol, one has $M\gg1$ copies of the identical states in the final idler-return joint state $\hat{\rho}_h^E$ of Eq.~\eqref{rho_E}. 
We can therefore apply the QCB for multiple hypotheses~\cite{li2016discriminating,nussbaum2011asymptotic} to obtain the asymptotic error probability. Due to the symmetry of the problem, the error exponent of the multiary hypothesis testing problem is equal to that of discrimination between two three-mode zero-mean Gaussian states with the covariance matrices~(see Appendix B)
\begin{align}
&
{\mathbf{V}}_{12I}^{(1)}=
\left(
\begin{array}{cccc}
(2N_B+1) {\mathbf I}_2&\bm 0&2\sqrt{\kappa}C_p{\mathbf Z}_2\\
\bm 0&(2N_B+1) {\mathbf I}_2&\bm 0\\
2\sqrt{\kappa}C_p{\mathbf Z}_2&\bm 0&(2N_S+1){\mathbf I}_2
\end{array} 
\right),
\nonumber
&
\\
&
{\mathbf{V}}_{12I}^{(2)}=
\left(
\begin{array}{cccc}
(2N_B+1) {\mathbf I}_2&\bm 0&\bm 0\\
\bm 0&(2N_B+1) {\mathbf I}_2&2\sqrt{\kappa}C_p{\mathbf Z}_2\\
\bm 0&2\sqrt{\kappa}C_p{\mathbf Z}_2&(2N_S+1){\mathbf I}_2
\end{array} 
\right).
\label{noisy_cov_3mode}
&
\end{align}
The error exponent can be analytically calculated~\cite{Pirandola2008}, leading to the asymptotic formula for the Helstrom limit when $N_B\gg1, N_S\ll1$ and $M\gg1$  as
\be 
P_{E,H} \sim \frac{m-1}{m}\exp\left[-\frac{2M\kappa N_S}{N_B}\right], 
\label{P_E_H_QCB}
\ee 
\QZ{which is tight in the error exponent. Here the constant $(m-1)/m$ is chosen to match the low signal-to-noise ratio limit with random guess}. Comparing with the optimal classical performance in Eq.~\eqref{P_C_H_QCB}, we see the EA case in Eq.~\eqref{P_E_H_QCB} has a factor of four (6dB) advantage in the error exponent, analog to the entanglement benefit in QI.

We can also derive a PGM~\cite{PGM1,PGM2,PGM3,zhuang2020entanglement}-based upper bound for the Helstrom limit~(see Appendix B)
\begin{align}
 P_{E,H} \le P_{E,UB}& = (m-1)F^M\left({\mathbf{V}}_{12I}^{(1)},{\mathbf{V}}_{12I}^{(2)}\right)
 \\
 &\simeq (m-1)\exp\left(-\frac{M\kappa N_S}{N_B}\right),
\label{UB2}
\end{align}
where $F\left(\bm V_1,\bm V_2\right)$ is the fidelity between two zero-mean Gaussian states with the covariance matrices $\bm V_1$ and $\bm V_2$~\cite{banchi2015}.
Indeed, we see the error-exponent of $P_{E,UB}$ is a factor of two worse than that of $P_{E,H}$. However, compared with the classical performances in Eqs.~\eqref{P_C_H_QCB} and~\eqref{P_CI_DD}, we still see a factor of 2 (3dB) advantage in the error exponent.

In Fig.~\ref{fig:m}, we confirm the advantage from entanglement. The entangled upper bound $P_{E,UB}$ (red dashed) offers rigorous advantages, as well as a scaling advantage in the error exponent, while the asymptotic performance $P_{E,H}$ (orange dashed) provides further advantages (the full expression is utilized for evaluation). \QZ{The QCB results (orange dashed for the entangled case and green dashed for the classical case) are tight in the error exponent, showing a rigorous 6dB advantage from entanglement. However, these bounds can be non-tight up to a constant factor independent of $M$, and thus do not show exact amount of advantages.} These results confirm the quantum advantage of entanglement in the ranging task, assuming an optimal receiver that jointly measures the entire quantum system of the collected light and the idler. \QZ{Note that such advantages only exist in the $N_B\gg1$ limit, and disappears when the noise is small.}

\begin{figure}[t]
    \centering
    \includegraphics[width=0.475\textwidth]{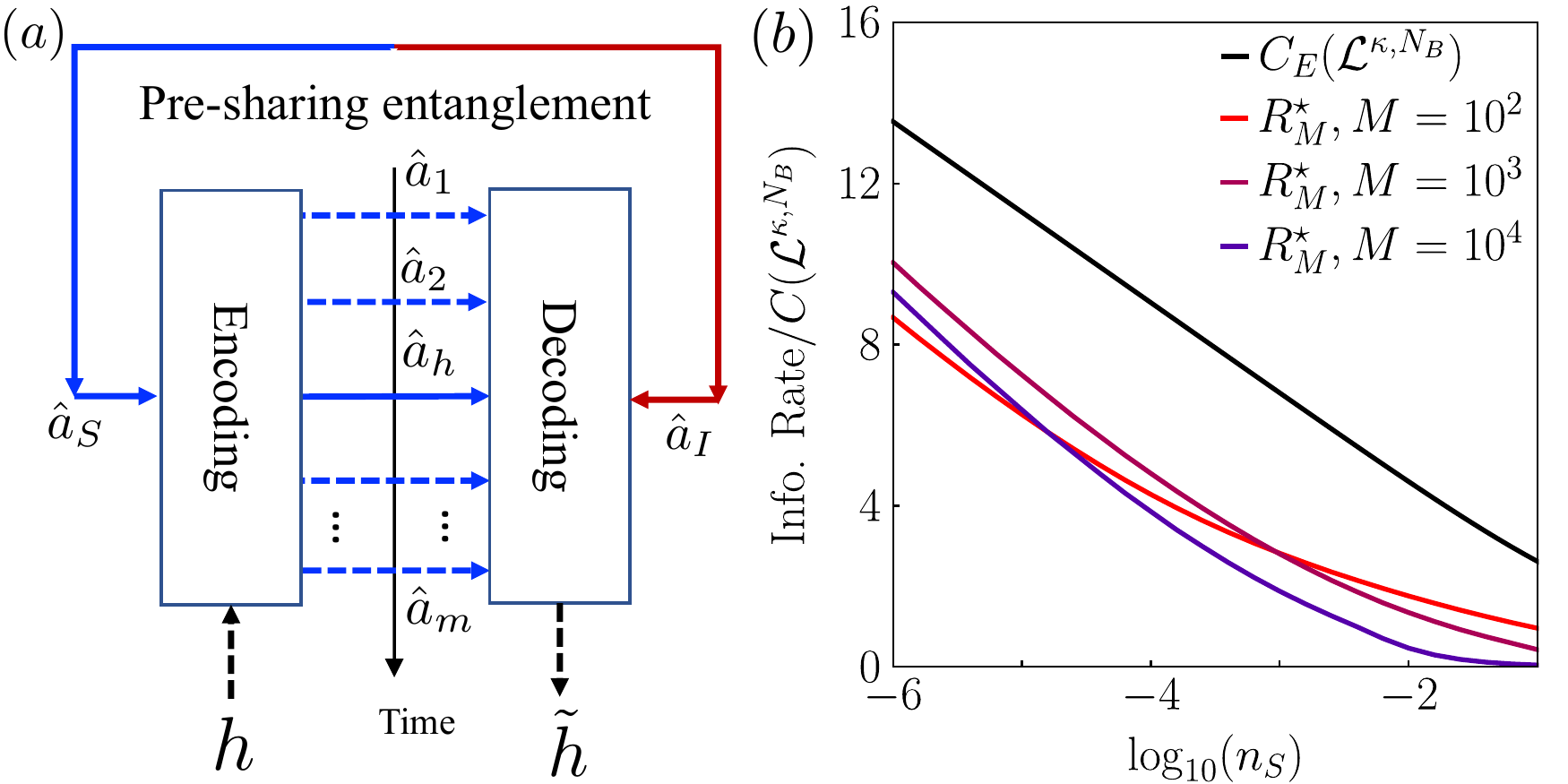}
    \caption{
    (a) Schematic of the entanglement-assisted communication protocol. (b) Information rate of the entanglement-assisted communication protocol, with $\kappa=0.1$ and $N_B=20$. We compare the optimized rates $R^\star_M$ for the fixed number of repetition modes $M=10^2, 10^3, 10^4$ (red, purple, blue) and the entanglement-assisted classical capacity (black), versus the signal average brightness $n_S$.
    \label{fig:schematic_EA_com}
    }
\end{figure}

{\em Entanglement-assisted communication.---} Our quantum ranging results can be applied to the design of pulse-position modulated EA communication, where entanglement is pre-shared between a sender and a receiver. As shown in Fig.~\ref{fig:schematic_EA_com}(a), to send the classical message $h\in[1,m]$, the sender chooses $m$ possible time slices to send the signal part $\hat{a}_S$ of the entangled TMSV to the receiver, who collects light continuously to obtain all modes $\{\hat{a}_\ell\}_{\ell=1}^m$ corresponding to the $m$ time slices. The receiver then decodes the classical message $\tilde{h}$ by determining which time slice contains the signal from the sender, via measuring the collected modes $\{\hat{a}_\ell\}_{\ell=1}^m$ jointly with the idler $\hat{a}_I$. 

In the ranging protocol of Fig.~\ref{fig:schematic}, suppose we put all the loss and noise to the receiver side, then the target's range can be considered as the modulation device of the sender, and the path from the observer to the target as the ideal noiseless channel for entanglement pre-sharing~(see Appendix D). The same result of Eq.~\eqref{P_E_H_QCB} gives the asymptotic optimal decoding error probability, leading to an information rate per mode as $R_{m,M}=I\left(P_{E,H}\right)/Mm$, where the mutual information
\be  
I\left(p\right)=\log_2\left(m\right)+\left[\left(1-p\right)\log_2\left(1-p\right)+p\log_2\left(\frac{p}{m-1}\right) \right].
\ee 
We choose the signal total mean photon number $N_S=mn_S$, giving $n_S$ photons being sent per mode \QZ{per time slice} on average. To achieve the best rate, we optimize over the number of time slices $m$ to obtain the optimal rate of EA communication
$
R_{M}^\star=\max_{m} R_{m,M}.
$ 

As benchmarks, we calculate the corresponding classical capacity~\cite{hausladen1996classical,schumacher1997sending,holevo1998capacity,giovannetti2014ultimate} $C(\mathcal{L}^{\kappa,N_B})$, with the mean photon number constrained to $n_S$. As Eq.~\eqref{P_E_H_QCB} is asymptotically tight, we consider $M\gg1$ and plot the ratio of information rate over $C(\mathcal{L}^{\kappa,N_B})$ in Fig.~\ref{fig:schematic_EA_com}(b) and indeed see a great advantage in the low brightness region. In fact, when compared with the EA capacity $C_E(\mathcal{L}^{\kappa,N_B})$ (black solid) that upper bounds all possible EA communication rates, we see that the rate $R_{M}^\star$ has the scaling $R_{M}^\star/C(\mathcal{L}^{\kappa,N_B})\sim \ln(1/N_S)$ versus the signal power, identical to the scaling of $C_E(\mathcal{L}^{\kappa,N_B})$~\cite{shi2020practical}. Therefore, the receiver design for the ranging protocol would also be able to offer a great advantage in EA communication \QZ{in the low rate region}. 

{\em Receiver design.---} Here we provide a practical receiver design for the ranging problem when $m=2$, based on the optical-parametric amplifier (OPA)~\cite{Guha2009} (see Appendix C). In this case of binary range discrimination, there are two groups of collected modes $\{\hat{a}_1^{(n)}\}_{n=1}^M$ and $\{\hat{a}_2^{(n)}\}_{n=1}^M$ corresponding to the two time slices. One can perform a phase shift on each block of modes and then perform a joint Gaussian operation with the idler modes to obtain
\be 
\hat{a}_I^{(n)\prime} =\sqrt{G}\hat{a}_I^{(n)} + \sqrt{\frac{G-1}{2}}\sum_{\ell=1}^2 e^{i\ell \pi}\hat{a}_\ell^{(n)\dagger}.
\ee  
To determine the target's range, we measure the total photon number of $\{\hat{a}_I^{(n)\prime}\}_{n=1}^M$, with each mode's mean photon number
$ 
\braket{\hat{a}_I^{(n)\prime\dagger} \hat{a}_I^{(n)\prime}}=G N_S+(G-1)(N_B+1)+2(-1)^hC_p\sqrt{{G(G-1)\kappa}/{2}} 
$
conditioned on hypothesis $h$. Therefore, the hypothesis can be determined from a threshold decision of the photon count. Choosing the optimal gain $G\sim1+2\sqrt{N_S}/N_B$, the error probability performance
$
P_{E,OPA}\simeq \exp\left[{-M\kappa N_S/N_B}\right]/2,
$
when $N_B\gg1, N_S\ll1$
providing a factor of two (3dB) advantage in the error exponent over the classical limit in Eq.~\eqref{P_C_H_QCB}. In Fig.~\ref{fig:m}(a), we plot the receiver performance (red solid), confirming the error exponent advantage. 

{\em Discussions.---} We propose a quantum ranging protocol enabled by entanglement to provide a 6dB advantage in the error exponent of determining the range among an arbitrary number of possibilities. To enable rigorous analyses, we have formulated the ranging problem as a hypothesis testing problem; the parameter estimation version would require a continuous-time treatment, which we defer to future works. The receiver design in the general case is an open problem. One potential approach is to design a non-demolition version of the sum-frequency-generation receiver design~\cite{zhuang2017entanglement}. The intuition is that the non-demolition measurement will allow one to utilize the same idler to interact with all collected modes until the correlated mode is located. 

\begin{acknowledgements}
Q.Z. acknowledges the Defense Advanced Research Projects Agency (DARPA) under Young Faculty Award (YFA) Grant No. N660012014029 and Craig M. Berge Dean's Faculty Fellowship of University of Arizona. Q.Z. thanks Saikat Guha, Stefano Pirandola and Haowei Shi for discussions. \QZ{Q.Z. acknowledges Jeffrey Shapiro for valuable feedback.}
\end{acknowledgements}

\appendix

\tableofcontents

\section{Two-mode squeezed vacuum}
In the entangled strategy, each signal-idler pair is in the TMSV described by the wave function
\be 
\ket{\phi^{\rm TMSV}}_{SI}=\sum_{n=0}^\infty \sqrt{\frac{N_{S}^{n}}{(N_{S}+1)^{n+1}}} \ket{n}_{S}\ket{n}_{I},
\label{eq:state_TMSV}
\ee 
where $\ket{n}$ is the number state defined by $\hat{a}^\dagger \hat{a}\ket{n}=n\ket{n}$. 

As zero-mean Gaussian states~\cite{Weedbrook2012}, a pair of TMSV is conveniently characterized by its covariance matrix
\begin{align}
&
{\mathbf{V}}_{\rm SI}=
\left(
\begin{array}{cccc}
(2N_S+1) {\mathbf I}_2&2C_p{\mathbf Z}_2\\
2C_p{\mathbf Z}_2&(2N_S+1){\mathbf I}_2
\end{array} 
\right),
\label{V_TMSS}
&
\end{align}
where $C_p=\sqrt{N_S\left(N_S+1\right)}$ and ${\mathbf Z}_2$ is the Pauli Z matrix.

\section{Quantum Chernoff bounds}
In the discrimination between states $\{\hat{\zeta}_h^{\otimes M} \}_{h=1}^m$ with an arbitrary prior, the asymptotic error exponent of the minimum error probability $P_H$ is given by the minimum pairwise error exponent~\cite{li2016discriminating,nussbaum2011asymptotic}, i.e.,
\be 
\lim_{M\to\infty} \frac{-1}{M} \ln\left(P_H\right) =\min_{h\neq h^\prime} C(\hat{\zeta}_h, \hat{\zeta}_{h^\prime})
\ee 
where the binary Chernoff exponent~\cite{audenaert2007discriminating}
\be 
C(\hat{\zeta}_h, \hat{\zeta}_{h^\prime})=\max_{0\le s \le 1} \left\{-\log \tr\left[ \hat{\zeta}_h^s \hat{\zeta}_{h^\prime}^{1-s}\right]\right\}.
\label{C_QCB}
\ee 
Note that $C$ is jointly convex~\cite{audenaert2007discriminating}.

\subsection{Asymptotic classical performance}
We consider all $M$-mode classical states with a positive P-function and show that single-mode coherent states minimizes the asymptotic error probability, in the sense that they maximizes the error exponent.

For the convenience of analysis, we will parameterize a coherent state $\ket{\alpha}$ with the phase and amplitude squared, i.e.,  $\ket{x,\theta}\equiv \ket{\sqrt{x}e^{i\theta}}$, where $x \ge 0$ and $0 \le \theta \le 2\pi$. In this notation, an $M$-mode coherent state $\ket{\bm x,\bm \theta}=\otimes_{k=1}^M \ket{x_k,\theta_k}$ is again a tensor product of multiple modes with generally-different amplitudes. Here $\bm x$ are positive and real vectors $\bm x=(x_1,\cdots, x_M)\equiv \{x_k\}_{k=1}^M$, and $\bm \theta=\{\theta_k\}_{k=1}^M$ is defined similarly.

In this notation, the general classical state
as the input can be written as a Lebesgue integral
\be
\hat{\rho}=\int{dP}  \ket{\bm x, \bm \theta}\bra{\bm x, \bm \theta},
\ee 
where the probability measure $P$ over $\bm x,\bm \theta$ can be arbitrary. Let us define 
\be
\|\bm x\|_1\equiv\sum_{k} |x_k|=\sum_{k} x_k,
\ee
which is the standard one-norm and equals the total mean photon number of the state $\ket{\bm x, \bm \theta}$. Then, the total energy constraint leads to the inequality
\be
\int{dP^\prime} \|\bm x\|_1 \le MN_S,
\label{energy_constraint_one_norm}
\ee 
where the integral has been simplified to a marginal probability measure $P^\prime$ restricted to the non-negative variables $\bm x$.

The total conditional state for hypothesis $h$ is also a mixture, with the expression
\begin{align}
\hat{\rho}_h^{\rm C}=\int{dP} \hat{\rho}^{\rm C}_{\bm x, \bm \theta,h},
\label{rho_C_P}
\end{align}
where each conditional state is given by
\be
\rho^{\rm C}_{\bm x, \bm \theta,h}=
\left(\otimes_{\ell\neq h} \hat{\sigma}^{(B)}_{\hat{\bm a}_{\ell}}\right)\otimes \hat{\sigma}^{(T)}_{\hat{\bm a}_{h}},
\ee
similar to Eq.~\eqref{rho_C} of the main paper.
The target state $\hat{\sigma}^{(T)}_{\hat{\bm a}_{h}}$ is a product of $M$ displaced thermal states, each with an amplitude $\sqrt{\kappa x_k}e^{i\theta_k}$ and a covariance matrix $(2N_B+1)\bm I$; the background state $\hat{\sigma}^{(B)}_{\hat{\bm a}_{\ell}}$ is a product of $M$ thermal states, each with zero mean and covariance matrix $(2N_B+1)\bm I$.

Given the same prior, the minimum error probability $P_H$ given by the Helstrom limit is lower bounded by the Helstrom limit of each component in Eq.~\eqref{rho_C_P} (see Lemma 2 of Ref.~\cite{zhuang2020entanglement}) 
\be 
P_H(\{\hat{\rho}_h^{\rm C}\}_{h=1}^m)\ge \int{dP} P_H\left(\{\hat{\rho}^{\rm C}_{\bm x, \bm \theta,h}\}_{h=1}^m\right).
\label{P_H_LB}
\ee 

Now we focus on each Helstrom limit $P_H\left(\{\hat{\rho}^{\rm C}_{\bm x, \bm \theta,h}\}_{h=1}^m\right)$. Consider a passive linear optics transform $\hat{U}_{\bm O}$ on the $M$ modes $\hat{\bm a}_{h}$ that makes the $M$ modes in the target state $\hat{\sigma}^{(T)}_{\hat{\bm a}_{h}}$ identical, i.e.
\be 
\hat{U}_{\bm O} \hat{\sigma}^{(T)}_{\hat{\bm a}_{h}}\hat{U}_{\bm O}^\dagger=\left(\hat{\sigma}^{(T)}_{h}\right)^{\otimes M},
\ee 
where effectively each copy $\hat{\sigma}^{(T)}_{h}=\calL_{\kappa,N_B}\left(\ket{\sqrt{\|\bm x\|_1/M}}\bra{\sqrt{\|\bm x\|_1/M}}\right)$.
Such a transform always exists, and we describe it by an orthogonal symplectic matrix $\bm O$. Consider applying the same passive linear optics transform on each block of modes $\hat{\bm a}_{\ell}$ for $1\le \ell \le m$, the background modes are still in the produc thermal state
\be 
\hat{U}_{\bm O}\hat{\sigma}^{(B)}_{\hat{\bm a}_{\ell}}\hat{U}_{\bm O}^\dagger =\hat{\sigma}^{(B)}_{\hat{\bm a}_{\ell}}=\left(\hat{\sigma}_\ell^{(B)}\right)^{\otimes M},
\ee 
where each copy $\hat{\sigma}_\ell^{(B)}$ is a thermal state with mean photon number $N_B$.

Such a transform will not change the Helstrom limit, 
\begin{align} 
&P_H\left(\{\hat{\rho}^{\rm C}_{\bm x, \bm \theta,h}\}_{h=1}^m\right)=P_H\left(\{\hat{U}_{\bm O}^{\otimes m}\hat{\rho}^{\rm C}_{\bm x, \bm \theta,h}\hat{U}_{\bm O}^{\otimes m^\dagger}\}_{h=1}^m\right)
\\
&=P_H\left(\{\left(\hat{\tilde{\rho}}^{\rm C}_{\bm x, \bm \theta,h}\right)^{\otimes M} \}_{h=1}^m\right),
\end{align} 
where
\be 
\hat{\tilde{\rho}}^{\rm C}_{\bm x, \bm \theta,h}=\hat{\sigma}^{(T)}_{h}\otimes  \left(\otimes_{\ell\neq h}\hat{\sigma}_\ell^{(B)}\right).
\label{rho_tensor}
\ee 
Now each state involved is a multiple copy of an identical state, and we can apply the asymptotically tight (when $M\gg 1$) QCB
\begin{align}
&\lim_{M\to\infty}\frac{-1}{M} \left(\ln P_H\left(\{\left(\hat{\tilde{\rho}}^{\rm C}_{\bm x, \bm \theta,h}\right)^{\otimes M} \}_{h=1}^m\right)\right)
\nonumber
\\
&=\min_{h\neq h^\prime} C\left(\hat{\tilde{\rho}}^{\rm C}_{\bm x, \bm \theta,h}, \hat{\tilde{\rho}}^{\rm C}_{\bm x, \bm \theta,h^\prime}\right)
\\
&=C\left(\hat{\tilde{\rho}}^{\rm C}_{\bm x, \bm \theta,1}, \hat{\tilde{\rho}}^{\rm C}_{\bm x, \bm \theta,2}\right)
\\
&=C\left(\hat{\sigma}^{(T)}_{1}\otimes \hat{\sigma}_2^{(B)}, \hat{\sigma}_1^{(B)}\otimes \hat{\sigma}^{(T)}_{2}\right)
\\
&=\frac{2\kappa \|\bm x\|_1/M}{1+2N_B+2\sqrt{N_B\left(1+N_B\right)}}.
\label{each_term_QCB}
\end{align} 
where we utilized the symmetry of the problem in the second equality, the form of state in Eq.~\eqref{rho_tensor} in the second last equality, and applied the analytical approach of evaluating the Chernoff exponent for Gaussian states~\cite{Pirandola2008} in the last equality.

Then we consider the error exponent of the overall task. From Ineq.~\eqref{P_H_LB} and the concavity of $\ln$ function, we have
\begin{align}
&\lim_{M\to\infty}\frac{-1}{M}\ln\left[P_H\left(\{\hat{\rho}_h^{\rm C}\}_{h=1}^m\right)\right]
\\
&\le 
\lim_{M\to\infty}
\frac{-1}{M}\int{dP} \ln\left[P_H\left(\{\hat{\rho}^{\rm C}_{\bm x, \bm \theta,h}\}_{h=1}^m\right)\right].
\\
&=\int{dP}  \frac{2\kappa \|\bm x\|_1/M}{1+2N_B+2\sqrt{N_B\left(1+N_B\right)}}
\\
&\le \frac{2\kappa N_S}{1+2N_B+2\sqrt{N_B\left(1+N_B\right)}}.
\end{align}
where in the first equality we utilized Eq.~\eqref{each_term_QCB} and in the last inequality we utlized the energy constraint in Ineq.~\eqref{energy_constraint_one_norm}. It is easy to verify that this upper bound is achieved by a coherent state input.

Therefore, we can have the optimal classical error probability of any classical input with a positive P-function as
\begin{align}
P_H\left(\{\hat{\rho}_h^{\rm C}\}_{h=1}^m\right)&\sim  \frac{m-1}{m}\exp\left[-\frac{2M\kappa N_S}{1+2N_B+2\sqrt{N_B\left(1+N_B\right)}}\right]
\\
&\simeq 
\frac{m-1}{m}\exp\left[-\frac{M\kappa N_S}{2N_B}\right].
\end{align}

\subsection{Asymptotic performance of the entangled scheme}

We reprint the joint state in Eq.~\eqref{rho_E} of the main paper
\begin{align}
\hat{\rho}_h^E&=\left(\otimes_{\ell\neq h} \hat{\sigma}^{(B)}_{\hat{\bm a}_{\ell}}\right)\otimes \hat{\Sigma}^{(T)}_{\hat{\bm a}_{h}\hat{\bm a}_I}=\left(\hat{\tilde{\rho}}_h^E\right)^{\otimes M}.
\label{rho_E_app}
\end{align}
where we have made the $M$-mode tensor structure explicit and 
\be 
\hat{\tilde{\rho}}_h^E=\left(\otimes_{\ell\neq h} \hat{\sigma}^{(B)}_\ell\right)\otimes \hat{\Sigma}^{(T)}_{hI}. 
\label{rho_E_structure}
\ee 
Here $\hat{\Sigma}^{(T)}_{hI}$ is a zero-mean Gaussian state with the covariance matrix ${\mathbf{V}}_{SI}^\prime$ in Eq.~\eqref{noisy_cov} of the main paper and $\hat{\sigma}^{(B)}$ is a thermal state with mean photon number $N_B$. 

As we have multiple copies of states, we can apply the QCB to obtain the asymptotic error probability. We can evaluate the Chernoff exponent in Eq.~\eqref{C_QCB}
\be 
C(\hat{\tilde{\rho}}_h^E, \hat{\tilde{\rho}}_{h^\prime}^E)=C(\hat{\tilde{\rho}}_1^E, \hat{\tilde{\rho}}_{2}^E)=C(\hat{\sigma}^{(B)}_1\otimes \hat{\Sigma}^{(T)}_{2I},\hat{\Sigma}^{(T)}_{1I}\otimes \hat{\sigma}^{(B)}_2),
\ee 
where we have utilized the symmetry and the structure of states in Eq.~\eqref{rho_E_structure}.
Now we arrive at the quantum Chernoff exponent between two three-mode zero-mean Gaussian states with covariance matrices in Eqs.~\eqref{noisy_cov_3mode} of the main paper. 
This can be analytically calculated via the approach in Ref.~\cite{Pirandola2008}, which leads to a lengthy formula. However, when $N_B\gg1, N_S\ll1$, we can have the leading order result as
\be 
C(\hat{\sigma}^{(B)}_1\otimes \hat{\Sigma}^{(T)}_{2I},\hat{\Sigma}^{(T)}_{1I}\otimes \hat{\sigma}^{(B)}_2)\simeq \frac{2\kappa N_S}{N_B}
\ee 

Therefore we have
\be 
\lim_{M\to\infty} \frac{-1}{M} \ln\left(P_H\left(\{\hat{\rho}_h^E\}\right)\right)\simeq \frac{2\kappa N_S}{N_B}.
\ee

\section{Practical receiver design}

\begin{figure}[b]
    \centering
    \includegraphics[width=0.45\textwidth]{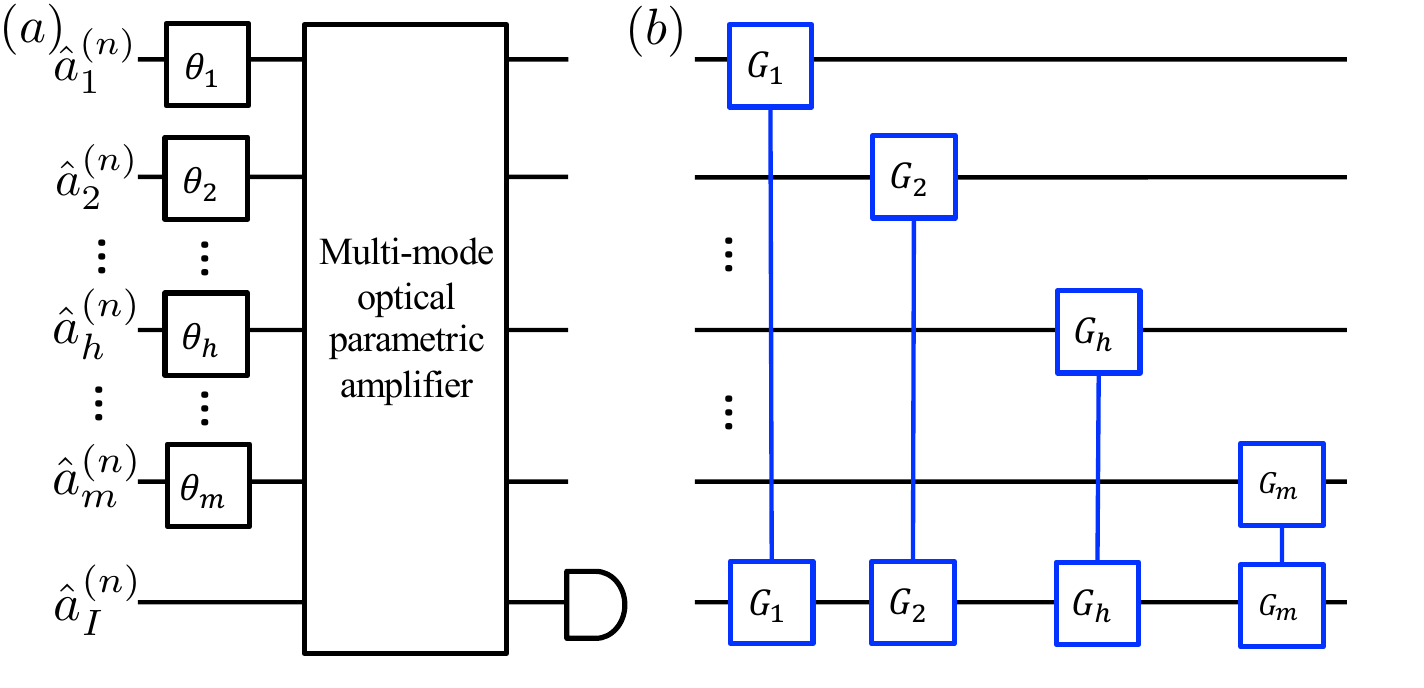}
    \caption{
    Schematic of the m-OPA receiver design.
    \label{fig:MOPA}
    }
\end{figure}

Below we address the problem of receiver design to realize the theoretical advantage. The receiver design for quantum ranging is much more challenging than that of QI, as the system now has $m+1$ subsystems, each composed of $M$ modes. We propose a receiver design based on a multi-mode optical parametric amplifier (m-OPA), which provides an advantage in the $m=2$ case. As shown in Fig.~\ref{fig:MOPA}(a), the receiver transforms each set of the phase-shifted $m$ received modes $\{\hat{a}^{(n)}_\ell\}_{\ell=1}^m$ and the stored idler to a new mode
\be 
\hat{a}_I^{(n)\prime} =\sqrt{G}\hat{a}_I^{(n)} + \sqrt{\frac{G-1}{m}}\sum_{\ell=1}^m e^{i\theta_\ell}\hat{a}_\ell^{(n)\dagger},
\ee 
where $\theta_\ell$ is the amount of phase shift prior to the m-OPA. We denote $\bm \theta=\{\theta_\ell\}_{\ell=1}^m$ for simplicity.
Then one performs photon counting on $\hat{a}_I^{(n)\prime}$ to obtain information about the hypothesis. 
As shown in Fig.~\ref{fig:MOPA}(b), the m-OPA can be completed via a sequence of ordinary OPAs, each implementing a two-mode squeezing as
\be  
\hat{a}_I^{(n)} \to \sqrt{G_\ell} \hat{a}_I^{(n)}+\sqrt{G_\ell-1} \hat{a}_\ell^{(n)\dagger}, 
\ee  
with the set of gains as the solution to the set of equations $\{(G_{\ell}-1)\prod_{k=\ell+1}^m G_k=(G-1)/m\}_{\ell=1}^m$.  
Under the above transform, the mean photon number conditioned on hypothesis $h$ 
\begin{align} 
\overline{N}(\theta_h)=\braket{\hat{a}_I^{(n)\prime\dagger} \hat{a}_I^{(n)\prime}}=&G N_S+(G-1)(N_B+1)
\nonumber
\\
&+2\sqrt{{G(G-1)\kappa}/{m}} \cos\theta_h C_p.
\end{align} %where we have utilized $\braket{\hat{a}_I \hat{a}_\ell}=0, \forall \ell\neq h$ and $\braket{\hat{a}_I \hat{a}_h}=\sqrt{\kappa}C_p$. 
We choose the gain $G=1+m\sqrt{N_S}/N_B$.

To begin with, one applies the same operations for the $M$ copies of the input and measure the total photon number. The distribution of the measurement outcome can be analytically solved~\cite{shi2020practical} as
\be 
P_{\rm OPA}(n|h)=\binom {n\!\!+\!\!M\!\!-\!\!1}{n}\!\left(\frac{\overline N(\theta_h)}{1+\overline N(\theta_h)}\right)^{\!n}\!\!\left(\frac{1}{1+\overline N(\theta_h)}\right)^{\!M},
\label{Pn_OPA}
\ee 
with the mean value $M\overline N(\theta_h)$ and the variance $M\overline \sigma^2(\theta_h)$, where we denote $\overline \sigma(\theta_h)=\sqrt{\overline N(\theta_h)\left(\overline N(\theta_h)+1\right)}$.

In the case of $m=2$, one can simply choose $\bm \theta=(0,\pi)$ and apply maximum-likelihood decision. When $M\gg1$, one can use Gaussian approximation to obtain the error probability
\be 
P_{E,OPA}=\frac{1}{2}{\rm Erfc}\left[\frac{M}{2}\frac{\overline N(0)-\overline N(\pi)}{\overline \sigma(0)+\overline \sigma(\pi)}\right]\simeq \frac{1}{2}e^{-M\kappa N_S/N_B}.
\ee 
We see for $m=2$, when $N_B\gg1, N_S\ll1$ the m-OPA receiver provides a factor of two (3dB) advantage over the classical limit in Eq.~\eqref{P_C_H_QCB} of the main paper and the direct detection scheme in Eq.~\eqref{P_CI_DD} of the main paper. In Fig.~\ref{fig:m} of the main paper, we confirm the scaling advantage. 

For $m\ge 3$, we do not have the receiver design yet. However, it is possible that a well-designed adaptive protocol based on tuning the phases $\bm \theta$ can lead to a similar advantage as in the $m=2$ case.

\section{Entanglement-assisted communication} 
In a pulse-position-modulated communication scenario, the sender encodes $\log_2(m)$ classical bits by sending out the input mode $\hat{a}_S$ during $m$ possible time slices. The receiver collects all the light during the $m$ possible time slices and aims to determine which slice contains the input signal. An idler is pre-shared to the receiver and can be used for the joint decoding.

The channel between the sender and receiver is $\calL_{\kappa,N_B}$. Therefore, with the input mixed in, the output mode
\be 
\hat{a}_h=\sqrt{\kappa}\hat{a}_S+\sqrt{1-\kappa}\hat{e}_h,
\ee 
while all other output modes only contain noise. The state at the receiver side has the same form with Eq.~\eqref{rho_E} of the main paper. The receiver now faces the same state discrimination task as the quantum ranging case.

\begin{widetext}
To obtain more accurate results, we recover the passive signatures in the signal brightness, and the three-mode zero-mean Gaussian states of interest now has the covariance matrices (similar to Eqs.~\eqref{noisy_cov_3mode})
\begin{align}
&
{\mathbf{V}}_{12I}^{(1)}=
\left(
\begin{array}{cccc}
(2(N_B+\kappa N_S)+1) {\mathbf I}_2&\bm 0&2\sqrt{\kappa}C_p{\mathbf Z}_2\\
\bm 0&(2N_B+1) {\mathbf I}_2&\bm 0\\
2\sqrt{\kappa}C_p{\mathbf Z}_2&\bm 0&(2N_S+1){\mathbf I}_2
\end{array} 
\right).
\label{noisy_cov_3mode_1}
&
\end{align}
and the other one obtained by switching the first two modes
\begin{align}
&
{\mathbf{V}}_{12I}^{(2)}=
\left(
\begin{array}{cccc}
(2N_B+1) {\mathbf I}_2&\bm 0&\bm 0\\
\bm 0&(2(N_B+\kappa N_S)+1) {\mathbf I}_2&2\sqrt{\kappa}C_p{\mathbf Z}_2\\
\bm 0&2\sqrt{\kappa}C_p{\mathbf Z}_2&(2N_S+1){\mathbf I}_2
\end{array} 
\right).
\label{noisy_cov_3mode_2}
&
\end{align}
\end{widetext}
The error exponent can be analytically calculated via the approach in Ref.~\cite{Pirandola2008}, which leads to a lengthy formula. When $N_B\gg1, N_S\ll1$, the result is identical to Eq.~\eqref{P_E_H_QCB} of the main paper.

The mutual information of a $m$-pulse-position modulation with a symmetric error probability $p$ can be obtained as
\be 
I\left(p\right)=\log_2\left(m\right)+\left[\left(1-p\right)\log_2\left(1-p\right)+p\log_2\left(\frac{p}{m-1}\right) \right].
\ee 
Therefore with $M$ modes and $m$ slices, given signal total mean photon number $MN_S=Mmn_S$, the rate (transmitted information per channel use per mode) is 
\be 
R_{m,M}=\frac{1}{Mm}I\left(P_{E,H}\right),
\label{EA_PPM}
\ee  
where $P_{E,H}$ is asymptotically 
\be 
P_{E,H} \sim \frac{m-1}{m}\exp\left[-\frac{2M\kappa m n_S}{N_B}\right],
\label{P_E_H_QCB_app}
\ee 
as given in Eq.~\eqref{P_E_H_QCB} of the main paper. To achieve the best rate, we optimize over the number of time slices $m$ to obtain
\be 
R_{M}^\star=\max_{m} R_{m,M}.
\ee 
Besides numerically evaluating the above expression, we can also perform asymptotic analyses. Choosing $m=1/(2M\kappa n_S/N_B)\gg1$, we have to leading order
\be 
R_{m,M}\sim  \frac{\kappa n_S \ln(n_S)}{N_B}.
\label{Rsym}
\ee

The classical limit of reliable communication rates is given by the Holevo-Schumacher-Westmoreland (HSW) classical capacity~\cite{hausladen1996classical,schumacher1997sending,holevo1998capacity}, solvable for the thermal-loss channel~\cite{giovannetti2014ultimate}
\be 
C(\mathcal{L}^{\kappa,N_B})=g\left(\kappa n_S+N_B\right)-g\left(N_B\right).
\ee 
where $g(n)=(n+1)\log_2(n+1)-n\log_2 n$ is the entropy of a thermal state with mean photon number $n$. The EA classical capacity is~\cite{bennett2002entanglement}
\be
C_E(\mathcal{L}^{\kappa,N_B})=g(n_S)+g(n_S^\prime)-g(A_+)-g(A_-),
\label{CE_formula}
\ee 
where 
$A_\pm=(D-1\pm(n_S^\prime-n_S))/2$, $n_S^\prime=\kappa n_S+N_B$ and $D=\sqrt{(n_S+n_S^\prime+1)^2-4\kappa n_S(n_S+1)}$. 
In the limit of $n_S\ll1, N_B\gg1$, we have
\begin{align}
&C(\mathcal{L}^{\kappa,N_B})\simeq \frac{\kappa n_S}{\ln(2)N_B},
\\
&C_E(\mathcal{L}^{\kappa,N_B}) \simeq \frac{\kappa n_S\ln(n_S)}{\ln(2)N_B}.
\end{align}

\QZ{
\section{Noiseless case}}

\QZ{
Consider the $N_B=0, \kappa=1$ case first. In this case, assume coherent state input and TMSV input for the classical and quantum cases, the Helstrom limit can be analytically solved due to the geometric uniform symmetry (GUS)~\cite{cariolaro2010theory} and the fact that the relevant states are all pure. For $\zeta=\braket{\psi_h|\psi_{h^\prime \neq h}}$, the Helstrom limit
\begin{align}
&P_H(m,\zeta)=\frac{m-1}{m^2}\left[\sqrt{1+(m-1)\zeta}-\sqrt{1-\zeta}\right]^2,
\label{P_H_PPM}
\end{align}
which is achievable by the `pretty good' measurement~\cite{PGM1,PGM2,PGM3}. In particular, note that for $m\zeta\ll1$ we have the asymptotic expansion
\be
P_H=\frac{1}{4}(m-1)\zeta^2+O(m^2\zeta^3).
\label{P_H_asym}
\ee
For the quantum case
\be 
\zeta_{E}=\left(\frac{1}{1+N_S}\right)^M \simeq \exp\left(-MN_S\right),
\ee 
while for the classical case
\be 
\zeta_C=\exp\left(-MN_S\right).
\ee 
Therefore, we do not see any advantage with entanglement.
We therefore does not expect any advantage when $N_B\ll1, \kappa\sim 1$.
}

%\bibliography{myref.bib}

\begin{thebibliography}{58}%
\makeatletter
\providecommand \@ifxundefined [1]{%
 \@ifx{#1\undefined}
}%
\providecommand \@ifnum [1]{%
 \ifnum #1\expandafter \@firstoftwo
 \else \expandafter \@secondoftwo
 \fi
}%
\providecommand \@ifx [1]{%
 \ifx #1\expandafter \@firstoftwo
 \else \expandafter \@secondoftwo
 \fi
}%
\providecommand \natexlab [1]{#1}%
\providecommand \enquote  [1]{``#1''}%
\providecommand \bibnamefont  [1]{#1}%
\providecommand \bibfnamefont [1]{#1}%
\providecommand \citenamefont [1]{#1}%
\providecommand \href@noop [0]{\@secondoftwo}%
\providecommand \href [0]{\begingroup \@sanitize@url \@href}%
\providecommand \@href[1]{\@@startlink{#1}\@@href}%
\providecommand \@@href[1]{\endgroup#1\@@endlink}%
\providecommand \@sanitize@url [0]{\catcode `\\12\catcode `\$12\catcode
  `\&12\catcode `\#12\catcode `\^12\catcode `\_12\catcode `\%12\relax}%
\providecommand \@@startlink[1]{}%
\providecommand \@@endlink[0]{}%
\providecommand \url  [0]{\begingroup\@sanitize@url \@url }%
\providecommand \@url [1]{\endgroup\@href {#1}{\urlprefix }}%
\providecommand \urlprefix  [0]{URL }%
\providecommand \Eprint [0]{\href }%
\providecommand \doibase [0]{https://doi.org/}%
\providecommand \selectlanguage [0]{\@gobble}%
\providecommand \bibinfo  [0]{\@secondoftwo}%
\providecommand \bibfield  [0]{\@secondoftwo}%
\providecommand \translation [1]{[#1]}%
\providecommand \BibitemOpen [0]{}%
\providecommand \bibitemStop [0]{}%
\providecommand \bibitemNoStop [0]{.\EOS\space}%
\providecommand \EOS [0]{\spacefactor3000\relax}%
\providecommand \BibitemShut  [1]{\csname bibitem#1\endcsname}%
\let\auto@bib@innerbib\@empty
%</preamble>
\bibitem [{\citenamefont {Gisin}\ \emph {et~al.}(2002)\citenamefont {Gisin},
  \citenamefont {Ribordy}, \citenamefont {Tittel},\ and\ \citenamefont
  {Zbinden}}]{gisin2002}%
  \BibitemOpen
  \bibfield  {author} {\bibinfo {author} {\bibfnamefont {N.}~\bibnamefont
  {Gisin}}, \bibinfo {author} {\bibfnamefont {G.}~\bibnamefont {Ribordy}},
  \bibinfo {author} {\bibfnamefont {W.}~\bibnamefont {Tittel}},\ and\ \bibinfo
  {author} {\bibfnamefont {H.}~\bibnamefont {Zbinden}},\ }\bibfield  {title}
  {\bibinfo {title} {Quantum cryptography},\ }\href
  {https://doi.org/10.1103/RevModPhys.74.145} {\bibfield  {journal} {\bibinfo
  {journal} {Rev. Mod. Phys.}\ }\textbf {\bibinfo {volume} {74}},\ \bibinfo
  {pages} {145} (\bibinfo {year} {2002})}\BibitemShut {NoStop}%
\bibitem [{\citenamefont {Xu}\ \emph {et~al.}(2020)\citenamefont {Xu},
  \citenamefont {Ma}, \citenamefont {Zhang}, \citenamefont {Lo},\ and\
  \citenamefont {Pan}}]{xu2020}%
  \BibitemOpen
  \bibfield  {author} {\bibinfo {author} {\bibfnamefont {F.}~\bibnamefont
  {Xu}}, \bibinfo {author} {\bibfnamefont {X.}~\bibnamefont {Ma}}, \bibinfo
  {author} {\bibfnamefont {Q.}~\bibnamefont {Zhang}}, \bibinfo {author}
  {\bibfnamefont {H.-K.}\ \bibnamefont {Lo}},\ and\ \bibinfo {author}
  {\bibfnamefont {J.-W.}\ \bibnamefont {Pan}},\ }\bibfield  {title} {\bibinfo
  {title} {Secure quantum key distribution with realistic devices},\ }\href
  {https://doi.org/10.1103/RevModPhys.92.025002} {\bibfield  {journal}
  {\bibinfo  {journal} {Rev. Mod. Phys.}\ }\textbf {\bibinfo {volume} {92}},\
  \bibinfo {pages} {025002} (\bibinfo {year} {2020})}\BibitemShut {NoStop}%
\bibitem [{\citenamefont {Pirandola}\ \emph {et~al.}(2020)\citenamefont
  {Pirandola}, \citenamefont {Andersen}, \citenamefont {Banchi}, \citenamefont
  {Berta}, \citenamefont {Bunandar}, \citenamefont {Colbeck}, \citenamefont
  {Englund}, \citenamefont {Gehring}, \citenamefont {Lupo}, \citenamefont
  {Ottaviani} \emph {et~al.}}]{pirandola2020advances}%
  \BibitemOpen
  \bibfield  {author} {\bibinfo {author} {\bibfnamefont {S.}~\bibnamefont
  {Pirandola}}, \bibinfo {author} {\bibfnamefont {U.~L.}\ \bibnamefont
  {Andersen}}, \bibinfo {author} {\bibfnamefont {L.}~\bibnamefont {Banchi}},
  \bibinfo {author} {\bibfnamefont {M.}~\bibnamefont {Berta}}, \bibinfo
  {author} {\bibfnamefont {D.}~\bibnamefont {Bunandar}}, \bibinfo {author}
  {\bibfnamefont {R.}~\bibnamefont {Colbeck}}, \bibinfo {author} {\bibfnamefont
  {D.}~\bibnamefont {Englund}}, \bibinfo {author} {\bibfnamefont
  {T.}~\bibnamefont {Gehring}}, \bibinfo {author} {\bibfnamefont
  {C.}~\bibnamefont {Lupo}}, \bibinfo {author} {\bibfnamefont {C.}~\bibnamefont
  {Ottaviani}}, \emph {et~al.},\ }\bibfield  {title} {\bibinfo {title}
  {Advances in quantum cryptography},\ }\href@noop {} {\bibfield  {journal}
  {\bibinfo  {journal} {Adv. Opt. Photonics}\ }\textbf {\bibinfo {volume}
  {12}},\ \bibinfo {pages} {1012} (\bibinfo {year} {2020})}\BibitemShut
  {NoStop}%
\bibitem [{\citenamefont {Preskill}(2018)}]{Preskill2018quantumcomputingin}%
  \BibitemOpen
  \bibfield  {author} {\bibinfo {author} {\bibfnamefont {J.}~\bibnamefont
  {Preskill}},\ }\bibfield  {title} {\bibinfo {title} {Quantum computing in the
  nisq era and beyond},\ }\href {https://doi.org/10.22331/q-2018-08-06-79}
  {\bibfield  {journal} {\bibinfo  {journal} {{Quantum}}\ }\textbf {\bibinfo
  {volume} {2}},\ \bibinfo {pages} {79} (\bibinfo {year} {2018})}\BibitemShut
  {NoStop}%
\bibitem [{\citenamefont {Giovannetti}\ \emph {et~al.}(2011)\citenamefont
  {Giovannetti}, \citenamefont {Lloyd},\ and\ \citenamefont
  {Maccone}}]{giovannetti2011advances}%
  \BibitemOpen
  \bibfield  {author} {\bibinfo {author} {\bibfnamefont {V.}~\bibnamefont
  {Giovannetti}}, \bibinfo {author} {\bibfnamefont {S.}~\bibnamefont {Lloyd}},\
  and\ \bibinfo {author} {\bibfnamefont {L.}~\bibnamefont {Maccone}},\
  }\bibfield  {title} {\bibinfo {title} {Advances in quantum metrology},\
  }\href@noop {} {\bibfield  {journal} {\bibinfo  {journal} {Nat. Photonics}\
  }\textbf {\bibinfo {volume} {5}},\ \bibinfo {pages} {222} (\bibinfo {year}
  {2011})}\BibitemShut {NoStop}%
\bibitem [{\citenamefont {Degen}\ \emph {et~al.}(2017)\citenamefont {Degen},
  \citenamefont {Reinhard},\ and\ \citenamefont
  {Cappellaro}}]{degen2017quantum}%
  \BibitemOpen
  \bibfield  {author} {\bibinfo {author} {\bibfnamefont {C.~L.}\ \bibnamefont
  {Degen}}, \bibinfo {author} {\bibfnamefont {F.}~\bibnamefont {Reinhard}},\
  and\ \bibinfo {author} {\bibfnamefont {P.}~\bibnamefont {Cappellaro}},\
  }\bibfield  {title} {\bibinfo {title} {Quantum sensing},\ }\href
  {https://doi.org/10.1103/RevModPhys.89.035002} {\bibfield  {journal}
  {\bibinfo  {journal} {Rev. Mod. Phys.}\ }\textbf {\bibinfo {volume} {89}},\
  \bibinfo {pages} {035002} (\bibinfo {year} {2017})}\BibitemShut {NoStop}%
\bibitem [{\citenamefont {Braun}\ \emph {et~al.}(2018)\citenamefont {Braun},
  \citenamefont {Adesso}, \citenamefont {Benatti}, \citenamefont {Floreanini},
  \citenamefont {Marzolino}, \citenamefont {Mitchell},\ and\ \citenamefont
  {Pirandola}}]{braun2018rmp}%
  \BibitemOpen
  \bibfield  {author} {\bibinfo {author} {\bibfnamefont {D.}~\bibnamefont
  {Braun}}, \bibinfo {author} {\bibfnamefont {G.}~\bibnamefont {Adesso}},
  \bibinfo {author} {\bibfnamefont {F.}~\bibnamefont {Benatti}}, \bibinfo
  {author} {\bibfnamefont {R.}~\bibnamefont {Floreanini}}, \bibinfo {author}
  {\bibfnamefont {U.}~\bibnamefont {Marzolino}}, \bibinfo {author}
  {\bibfnamefont {M.~W.}\ \bibnamefont {Mitchell}},\ and\ \bibinfo {author}
  {\bibfnamefont {S.}~\bibnamefont {Pirandola}},\ }\bibfield  {title} {\bibinfo
  {title} {Quantum-enhanced measurements without entanglement},\ }\href
  {https://doi.org/10.1103/RevModPhys.90.035006} {\bibfield  {journal}
  {\bibinfo  {journal} {Rev. Mod. Phys.}\ }\textbf {\bibinfo {volume} {90}},\
  \bibinfo {pages} {035006} (\bibinfo {year} {2018})}\BibitemShut {NoStop}%
\bibitem [{\citenamefont {Pirandola}\ \emph {et~al.}(2018)\citenamefont
  {Pirandola}, \citenamefont {Bardhan}, \citenamefont {Gehring}, \citenamefont
  {Weedbrook},\ and\ \citenamefont {Lloyd}}]{pirandola2018advances}%
  \BibitemOpen
  \bibfield  {author} {\bibinfo {author} {\bibfnamefont {S.}~\bibnamefont
  {Pirandola}}, \bibinfo {author} {\bibfnamefont {B.~R.}\ \bibnamefont
  {Bardhan}}, \bibinfo {author} {\bibfnamefont {T.}~\bibnamefont {Gehring}},
  \bibinfo {author} {\bibfnamefont {C.}~\bibnamefont {Weedbrook}},\ and\
  \bibinfo {author} {\bibfnamefont {S.}~\bibnamefont {Lloyd}},\ }\bibfield
  {title} {\bibinfo {title} {Advances in photonic quantum sensing},\
  }\href@noop {} {\bibfield  {journal} {\bibinfo  {journal} {Nat. Photonics}\
  }\textbf {\bibinfo {volume} {12}},\ \bibinfo {pages} {724} (\bibinfo {year}
  {2018})}\BibitemShut {NoStop}%
\bibitem [{\citenamefont {Sidhu}\ and\ \citenamefont
  {Kok}(2020)}]{sidhu2020geometric}%
  \BibitemOpen
  \bibfield  {author} {\bibinfo {author} {\bibfnamefont {J.~S.}\ \bibnamefont
  {Sidhu}}\ and\ \bibinfo {author} {\bibfnamefont {P.}~\bibnamefont {Kok}},\
  }\bibfield  {title} {\bibinfo {title} {Geometric perspective on quantum
  parameter estimation},\ }\href@noop {} {\bibfield  {journal} {\bibinfo
  {journal} {AVS Quantum Science}\ }\textbf {\bibinfo {volume} {2}},\ \bibinfo
  {pages} {014701} (\bibinfo {year} {2020})}\BibitemShut {NoStop}%
\bibitem [{\citenamefont {Shor}(1997)}]{Shor_1997}%
  \BibitemOpen
  \bibfield  {author} {\bibinfo {author} {\bibfnamefont {P.}~\bibnamefont
  {Shor}},\ }\bibfield  {title} {\bibinfo {title} {Polynomial-time algorithms
  for prime factorization and discrete logarithms on a quantum computer},\
  }\href {https://doi.org/10.1137/S0097539795293172} {\bibfield  {journal}
  {\bibinfo  {journal} {SIAM J. Comput.}\ }\textbf {\bibinfo {volume} {26}},\
  \bibinfo {pages} {1484} (\bibinfo {year} {1997})}\BibitemShut {NoStop}%
\bibitem [{\citenamefont {Bennett}\ and\ \citenamefont
  {Wiesner}(1992)}]{bennett1992}%
  \BibitemOpen
  \bibfield  {author} {\bibinfo {author} {\bibfnamefont {C.~H.}\ \bibnamefont
  {Bennett}}\ and\ \bibinfo {author} {\bibfnamefont {S.~J.}\ \bibnamefont
  {Wiesner}},\ }\bibfield  {title} {\bibinfo {title} {Communication via one-and
  two-particle operators on einstein-podolsky-rosen states},\ }\href@noop {}
  {\bibfield  {journal} {\bibinfo  {journal} {Phys. Rev. Lett.}\ }\textbf
  {\bibinfo {volume} {69}},\ \bibinfo {pages} {2881} (\bibinfo {year}
  {1992})}\BibitemShut {NoStop}%
\bibitem [{\citenamefont {Bennett}\ \emph {et~al.}(2002)\citenamefont
  {Bennett}, \citenamefont {Shor}, \citenamefont {Smolin},\ and\ \citenamefont
  {Thapliyal}}]{bennett2002entanglement}%
  \BibitemOpen
  \bibfield  {author} {\bibinfo {author} {\bibfnamefont {C.~H.}\ \bibnamefont
  {Bennett}}, \bibinfo {author} {\bibfnamefont {P.~W.}\ \bibnamefont {Shor}},
  \bibinfo {author} {\bibfnamefont {J.~A.}\ \bibnamefont {Smolin}},\ and\
  \bibinfo {author} {\bibfnamefont {A.~V.}\ \bibnamefont {Thapliyal}},\
  }\bibfield  {title} {\bibinfo {title} {Entanglement-assisted capacity of a
  quantum channel and the reverse shannon theorem},\ }\href@noop {} {\bibfield
  {journal} {\bibinfo  {journal} {IEEE Trans. Inf. Theory}\ }\textbf {\bibinfo
  {volume} {48}},\ \bibinfo {pages} {2637} (\bibinfo {year}
  {2002})}\BibitemShut {NoStop}%
\bibitem [{\citenamefont {Bennett}\ and\ \citenamefont
  {Brassard}(2014)}]{Bennett20147}%
  \BibitemOpen
  \bibfield  {author} {\bibinfo {author} {\bibfnamefont {C.~H.}\ \bibnamefont
  {Bennett}}\ and\ \bibinfo {author} {\bibfnamefont {G.}~\bibnamefont
  {Brassard}},\ }\bibfield  {title} {\bibinfo {title} {Quantum cryptography:
  Public key distribution and coin tossing},\ }\href
  {https://doi.org/http://dx.doi.org/10.1016/j.tcs.2014.05.025} {\bibfield
  {journal} {\bibinfo  {journal} {Theoretical Computer Science}\ }\textbf
  {\bibinfo {volume} {560, Part 1}},\ \bibinfo {pages} {7 } (\bibinfo {year}
  {2014})}\BibitemShut {NoStop}%
\bibitem [{\citenamefont {Ekert}(1991)}]{Ekert_1991}%
  \BibitemOpen
  \bibfield  {author} {\bibinfo {author} {\bibfnamefont {A.~K.}\ \bibnamefont
  {Ekert}},\ }\bibfield  {title} {\bibinfo {title} {Quantum cryptography based
  on bell's theorem},\ }\href {https://doi.org/10.1103/PhysRevLett.67.661}
  {\bibfield  {journal} {\bibinfo  {journal} {Phys. Rev. Lett.}\ }\textbf
  {\bibinfo {volume} {67}},\ \bibinfo {pages} {661} (\bibinfo {year}
  {1991})}\BibitemShut {NoStop}%
\bibitem [{\citenamefont {Zwierz}\ \emph {et~al.}(2010)\citenamefont {Zwierz},
  \citenamefont {P{\'e}rez-Delgado},\ and\ \citenamefont
  {Kok}}]{zwierz2010general}%
  \BibitemOpen
  \bibfield  {author} {\bibinfo {author} {\bibfnamefont {M.}~\bibnamefont
  {Zwierz}}, \bibinfo {author} {\bibfnamefont {C.~A.}\ \bibnamefont
  {P{\'e}rez-Delgado}},\ and\ \bibinfo {author} {\bibfnamefont
  {P.}~\bibnamefont {Kok}},\ }\bibfield  {title} {\bibinfo {title} {General
  optimality of the heisenberg limit for quantum metrology},\ }\href@noop {}
  {\bibfield  {journal} {\bibinfo  {journal} {Phys. Rev. Lett.}\ }\textbf
  {\bibinfo {volume} {105}},\ \bibinfo {pages} {180402} (\bibinfo {year}
  {2010})}\BibitemShut {NoStop}%
\bibitem [{\citenamefont {Giovannetti}\ \emph {et~al.}(2006)\citenamefont
  {Giovannetti}, \citenamefont {Lloyd},\ and\ \citenamefont
  {Maccone}}]{giovannetti2006}%
  \BibitemOpen
  \bibfield  {author} {\bibinfo {author} {\bibfnamefont {V.}~\bibnamefont
  {Giovannetti}}, \bibinfo {author} {\bibfnamefont {S.}~\bibnamefont {Lloyd}},\
  and\ \bibinfo {author} {\bibfnamefont {L.}~\bibnamefont {Maccone}},\
  }\bibfield  {title} {\bibinfo {title} {Quantum metrology},\ }\href@noop {}
  {\bibfield  {journal} {\bibinfo  {journal} {Phys. Rev. Lett.}\ }\textbf
  {\bibinfo {volume} {96}},\ \bibinfo {pages} {010401} (\bibinfo {year}
  {2006})}\BibitemShut {NoStop}%
\bibitem [{\citenamefont {Ge}\ \emph {et~al.}(2018)\citenamefont {Ge},
  \citenamefont {Jacobs}, \citenamefont {Eldredge}, \citenamefont {Gorshkov},\
  and\ \citenamefont {Foss-Feig}}]{ge2017distributed}%
  \BibitemOpen
  \bibfield  {author} {\bibinfo {author} {\bibfnamefont {W.}~\bibnamefont
  {Ge}}, \bibinfo {author} {\bibfnamefont {K.}~\bibnamefont {Jacobs}}, \bibinfo
  {author} {\bibfnamefont {Z.}~\bibnamefont {Eldredge}}, \bibinfo {author}
  {\bibfnamefont {A.~V.}\ \bibnamefont {Gorshkov}},\ and\ \bibinfo {author}
  {\bibfnamefont {M.}~\bibnamefont {Foss-Feig}},\ }\bibfield  {title} {\bibinfo
  {title} {Distributed quantum metrology with linear networks and separable
  inputs},\ }\href {https://doi.org/10.1103/PhysRevLett.121.043604} {\bibfield
  {journal} {\bibinfo  {journal} {Phys. Rev. Lett.}\ }\textbf {\bibinfo
  {volume} {121}},\ \bibinfo {pages} {043604} (\bibinfo {year}
  {2018})}\BibitemShut {NoStop}%
\bibitem [{\citenamefont {Proctor}\ \emph {et~al.}(2018)\citenamefont
  {Proctor}, \citenamefont {Knott},\ and\ \citenamefont
  {Dunningham}}]{proctor2017multi}%
  \BibitemOpen
  \bibfield  {author} {\bibinfo {author} {\bibfnamefont {T.~J.}\ \bibnamefont
  {Proctor}}, \bibinfo {author} {\bibfnamefont {P.~A.}\ \bibnamefont {Knott}},\
  and\ \bibinfo {author} {\bibfnamefont {J.~A.}\ \bibnamefont {Dunningham}},\
  }\bibfield  {title} {\bibinfo {title} {Multiparameter estimation in networked
  quantum sensors},\ }\href@noop {} {\bibfield  {journal} {\bibinfo  {journal}
  {Phys. Rev. Lett.}\ }\textbf {\bibinfo {volume} {120}},\ \bibinfo {pages}
  {080501} (\bibinfo {year} {2018})}\BibitemShut {NoStop}%
\bibitem [{\citenamefont {Zhuang}\ \emph {et~al.}(2018)\citenamefont {Zhuang},
  \citenamefont {Zhang},\ and\ \citenamefont
  {Shapiro}}]{zhuang2018distributed}%
  \BibitemOpen
  \bibfield  {author} {\bibinfo {author} {\bibfnamefont {Q.}~\bibnamefont
  {Zhuang}}, \bibinfo {author} {\bibfnamefont {Z.}~\bibnamefont {Zhang}},\ and\
  \bibinfo {author} {\bibfnamefont {J.~H.}\ \bibnamefont {Shapiro}},\
  }\bibfield  {title} {\bibinfo {title} {Distributed quantum sensing using
  continuous-variable multipartite entanglement},\ }\href@noop {} {\bibfield
  {journal} {\bibinfo  {journal} {Phys. Rev. A}\ }\textbf {\bibinfo {volume}
  {97}},\ \bibinfo {pages} {032329} (\bibinfo {year} {2018})}\BibitemShut
  {NoStop}%
\bibitem [{\citenamefont {Eldredge}\ \emph {et~al.}(2018)\citenamefont
  {Eldredge}, \citenamefont {Foss-Feig}, \citenamefont {Gross}, \citenamefont
  {Rolston},\ and\ \citenamefont {Gorshkov}}]{eldredge2018optimal}%
  \BibitemOpen
  \bibfield  {author} {\bibinfo {author} {\bibfnamefont {Z.}~\bibnamefont
  {Eldredge}}, \bibinfo {author} {\bibfnamefont {M.}~\bibnamefont {Foss-Feig}},
  \bibinfo {author} {\bibfnamefont {J.~A.}\ \bibnamefont {Gross}}, \bibinfo
  {author} {\bibfnamefont {S.~L.}\ \bibnamefont {Rolston}},\ and\ \bibinfo
  {author} {\bibfnamefont {A.~V.}\ \bibnamefont {Gorshkov}},\ }\bibfield
  {title} {\bibinfo {title} {Optimal and secure measurement protocols for
  quantum sensor networks},\ }\href@noop {} {\bibfield  {journal} {\bibinfo
  {journal} {Phys. Rev. A}\ }\textbf {\bibinfo {volume} {97}},\ \bibinfo
  {pages} {042337} (\bibinfo {year} {2018})}\BibitemShut {NoStop}%
\bibitem [{\citenamefont {Zhang}\ and\ \citenamefont
  {Zhuang}(2020)}]{zhang2020distributed}%
  \BibitemOpen
  \bibfield  {author} {\bibinfo {author} {\bibfnamefont {Z.}~\bibnamefont
  {Zhang}}\ and\ \bibinfo {author} {\bibfnamefont {Q.}~\bibnamefont {Zhuang}},\
  }\bibfield  {title} {\bibinfo {title} {Distributed quantum sensing},\
  }\href@noop {} {\bibfield  {journal} {\bibinfo  {journal} {Quantum Science
  and Technology}\ } (\bibinfo {year} {2020})}\BibitemShut {NoStop}%
\bibitem [{\citenamefont {Shi}\ \emph {et~al.}(2020)\citenamefont {Shi},
  \citenamefont {Zhang},\ and\ \citenamefont {Zhuang}}]{shi2020practical}%
  \BibitemOpen
  \bibfield  {author} {\bibinfo {author} {\bibfnamefont {H.}~\bibnamefont
  {Shi}}, \bibinfo {author} {\bibfnamefont {Z.}~\bibnamefont {Zhang}},\ and\
  \bibinfo {author} {\bibfnamefont {Q.}~\bibnamefont {Zhuang}},\ }\bibfield
  {title} {\bibinfo {title} {Practical route to entanglement-assisted
  communication over noisy bosonic channels},\ }\href
  {https://doi.org/10.1103/PhysRevApplied.13.034029} {\bibfield  {journal}
  {\bibinfo  {journal} {Phys. Rev. Applied}\ }\textbf {\bibinfo {volume}
  {13}},\ \bibinfo {pages} {034029} (\bibinfo {year} {2020})}\BibitemShut
  {NoStop}%
\bibitem [{\citenamefont {Hao}\ \emph {et~al.}(2020)\citenamefont {Hao},
  \citenamefont {Shi}, \citenamefont {Li}, \citenamefont {Zhuang},\ and\
  \citenamefont {Zhang}}]{hao2020}%
  \BibitemOpen
  \bibfield  {author} {\bibinfo {author} {\bibfnamefont {S.}~\bibnamefont
  {Hao}}, \bibinfo {author} {\bibfnamefont {H.}~\bibnamefont {Shi}}, \bibinfo
  {author} {\bibfnamefont {W.}~\bibnamefont {Li}}, \bibinfo {author}
  {\bibfnamefont {Q.}~\bibnamefont {Zhuang}},\ and\ \bibinfo {author}
  {\bibfnamefont {Z.}~\bibnamefont {Zhang}},\ }\bibfield  {title} {\bibinfo
  {title} {Entanglement-assisted communication surpassing the ultimate
  classical capacity},\ }\href@noop {} {\bibfield  {journal} {\bibinfo
  {journal} {submitted}\ } (\bibinfo {year} {2020})}\BibitemShut {NoStop}%
\bibitem [{\citenamefont {Lloyd}(2008)}]{Lloyd2008}%
  \BibitemOpen
  \bibfield  {author} {\bibinfo {author} {\bibfnamefont {S.}~\bibnamefont
  {Lloyd}},\ }\bibfield  {title} {\bibinfo {title} {Enhanced sensitivity of
  photodetection via quantum illumination},\ }\href
  {https://doi.org/10.1126/science.1160627} {\bibfield  {journal} {\bibinfo
  {journal} {Science}\ }\textbf {\bibinfo {volume} {321}},\ \bibinfo {pages}
  {1463} (\bibinfo {year} {2008})}\BibitemShut {NoStop}%
\bibitem [{\citenamefont {Tan}\ \emph {et~al.}(2008)\citenamefont {Tan},
  \citenamefont {Erkmen}, \citenamefont {Giovannetti}, \citenamefont {Guha},
  \citenamefont {Lloyd}, \citenamefont {Maccone}, \citenamefont {Pirandola},\
  and\ \citenamefont {Shapiro}}]{tan2008quantum}%
  \BibitemOpen
  \bibfield  {author} {\bibinfo {author} {\bibfnamefont {S.-H.}\ \bibnamefont
  {Tan}}, \bibinfo {author} {\bibfnamefont {B.~I.}\ \bibnamefont {Erkmen}},
  \bibinfo {author} {\bibfnamefont {V.}~\bibnamefont {Giovannetti}}, \bibinfo
  {author} {\bibfnamefont {S.}~\bibnamefont {Guha}}, \bibinfo {author}
  {\bibfnamefont {S.}~\bibnamefont {Lloyd}}, \bibinfo {author} {\bibfnamefont
  {L.}~\bibnamefont {Maccone}}, \bibinfo {author} {\bibfnamefont
  {S.}~\bibnamefont {Pirandola}},\ and\ \bibinfo {author} {\bibfnamefont
  {J.~H.}\ \bibnamefont {Shapiro}},\ }\bibfield  {title} {\bibinfo {title}
  {Quantum illumination with gaussian states},\ }\href@noop {} {\bibfield
  {journal} {\bibinfo  {journal} {Phys. Rev. Lett.}\ }\textbf {\bibinfo
  {volume} {101}},\ \bibinfo {pages} {253601} (\bibinfo {year}
  {2008})}\BibitemShut {NoStop}%
\bibitem [{\citenamefont {Guha}\ and\ \citenamefont {Erkmen}(2009)}]{Guha2009}%
  \BibitemOpen
  \bibfield  {author} {\bibinfo {author} {\bibfnamefont {S.}~\bibnamefont
  {Guha}}\ and\ \bibinfo {author} {\bibfnamefont {B.~I.}\ \bibnamefont
  {Erkmen}},\ }\bibfield  {title} {\bibinfo {title} {Gaussian-state
  quantum-illumination receivers for target detection},\ }\href
  {https://doi.org/10.1103/PhysRevA.80.052310} {\bibfield  {journal} {\bibinfo
  {journal} {Phys. Rev. A}\ }\textbf {\bibinfo {volume} {80}},\ \bibinfo
  {pages} {052310} (\bibinfo {year} {2009})}\BibitemShut {NoStop}%
\bibitem [{\citenamefont {Zhang}\ \emph {et~al.}(2013)\citenamefont {Zhang},
  \citenamefont {Tengner}, \citenamefont {Zhong}, \citenamefont {Wong},\ and\
  \citenamefont {Shapiro}}]{zhang2013}%
  \BibitemOpen
  \bibfield  {author} {\bibinfo {author} {\bibfnamefont {Z.}~\bibnamefont
  {Zhang}}, \bibinfo {author} {\bibfnamefont {M.}~\bibnamefont {Tengner}},
  \bibinfo {author} {\bibfnamefont {T.}~\bibnamefont {Zhong}}, \bibinfo
  {author} {\bibfnamefont {F.~N.~C.}\ \bibnamefont {Wong}},\ and\ \bibinfo
  {author} {\bibfnamefont {J.~H.}\ \bibnamefont {Shapiro}},\ }\bibfield
  {title} {\bibinfo {title} {Entanglement's benefit survives an
  entanglement-breaking channel},\ }\href
  {https://doi.org/10.1103/PhysRevLett.111.010501} {\bibfield  {journal}
  {\bibinfo  {journal} {Phys. Rev. Lett.}\ }\textbf {\bibinfo {volume} {111}},\
  \bibinfo {pages} {010501} (\bibinfo {year} {2013})}\BibitemShut {NoStop}%
\bibitem [{\citenamefont {Zhang}\ \emph {et~al.}(2015)\citenamefont {Zhang},
  \citenamefont {Mouradian}, \citenamefont {Wong},\ and\ \citenamefont
  {Shapiro}}]{zhang2015}%
  \BibitemOpen
  \bibfield  {author} {\bibinfo {author} {\bibfnamefont {Z.}~\bibnamefont
  {Zhang}}, \bibinfo {author} {\bibfnamefont {S.}~\bibnamefont {Mouradian}},
  \bibinfo {author} {\bibfnamefont {F.~N.~C.}\ \bibnamefont {Wong}},\ and\
  \bibinfo {author} {\bibfnamefont {J.~H.}\ \bibnamefont {Shapiro}},\
  }\bibfield  {title} {\bibinfo {title} {Entanglement-enhanced sensing in a
  lossy and noisy environment},\ }\href
  {https://doi.org/10.1103/PhysRevLett.114.110506} {\bibfield  {journal}
  {\bibinfo  {journal} {Phys. Rev. Lett.}\ }\textbf {\bibinfo {volume} {114}},\
  \bibinfo {pages} {110506} (\bibinfo {year} {2015})}\BibitemShut {NoStop}%
\bibitem [{\citenamefont {Lopaeva}\ \emph {et~al.}(2013)\citenamefont
  {Lopaeva}, \citenamefont {Ruo~Berchera}, \citenamefont {Degiovanni},
  \citenamefont {Olivares}, \citenamefont {Brida},\ and\ \citenamefont
  {Genovese}}]{Lopaeva_2013}%
  \BibitemOpen
  \bibfield  {author} {\bibinfo {author} {\bibfnamefont {E.~D.}\ \bibnamefont
  {Lopaeva}}, \bibinfo {author} {\bibfnamefont {I.}~\bibnamefont
  {Ruo~Berchera}}, \bibinfo {author} {\bibfnamefont {I.~P.}\ \bibnamefont
  {Degiovanni}}, \bibinfo {author} {\bibfnamefont {S.}~\bibnamefont
  {Olivares}}, \bibinfo {author} {\bibfnamefont {G.}~\bibnamefont {Brida}},\
  and\ \bibinfo {author} {\bibfnamefont {M.}~\bibnamefont {Genovese}},\
  }\bibfield  {title} {\bibinfo {title} {Experimental realization of quantum
  illumination},\ }\href {https://doi.org/10.1103/PhysRevLett.110.153603}
  {\bibfield  {journal} {\bibinfo  {journal} {Phys. Rev. Lett.}\ }\textbf
  {\bibinfo {volume} {110}},\ \bibinfo {pages} {153603} (\bibinfo {year}
  {2013})}\BibitemShut {NoStop}%
\bibitem [{\citenamefont {Zhuang}\ \emph
  {et~al.}(2017{\natexlab{a}})\citenamefont {Zhuang}, \citenamefont {Zhang},\
  and\ \citenamefont {Shapiro}}]{zhuang2017}%
  \BibitemOpen
  \bibfield  {author} {\bibinfo {author} {\bibfnamefont {Q.}~\bibnamefont
  {Zhuang}}, \bibinfo {author} {\bibfnamefont {Z.}~\bibnamefont {Zhang}},\ and\
  \bibinfo {author} {\bibfnamefont {J.~H.}\ \bibnamefont {Shapiro}},\
  }\bibfield  {title} {\bibinfo {title} {Optimum mixed-state discrimination for
  noisy entanglement-enhanced sensing},\ }\href
  {https://doi.org/10.1103/PhysRevLett.118.040801} {\bibfield  {journal}
  {\bibinfo  {journal} {Phys. Rev. Lett.}\ }\textbf {\bibinfo {volume} {118}},\
  \bibinfo {pages} {040801} (\bibinfo {year} {2017}{\natexlab{a}})}\BibitemShut
  {NoStop}%
\bibitem [{\citenamefont {Zhuang}\ \emph
  {et~al.}(2017{\natexlab{b}})\citenamefont {Zhuang}, \citenamefont {Zhang},\
  and\ \citenamefont {Shapiro}}]{zhuang2017NP}%
  \BibitemOpen
  \bibfield  {author} {\bibinfo {author} {\bibfnamefont {Q.}~\bibnamefont
  {Zhuang}}, \bibinfo {author} {\bibfnamefont {Z.}~\bibnamefont {Zhang}},\ and\
  \bibinfo {author} {\bibfnamefont {J.~H.}\ \bibnamefont {Shapiro}},\
  }\bibfield  {title} {\bibinfo {title} {Entanglement-enhanced neyman--pearson
  target detection using quantum illumination},\ }\href@noop {} {\bibfield
  {journal} {\bibinfo  {journal} {JOSA B}\ }\textbf {\bibinfo {volume} {34}},\
  \bibinfo {pages} {1567} (\bibinfo {year} {2017}{\natexlab{b}})}\BibitemShut
  {NoStop}%
\bibitem [{\citenamefont {Zhuang}\ \emph
  {et~al.}(2017{\natexlab{c}})\citenamefont {Zhuang}, \citenamefont {Zhang},\
  and\ \citenamefont {Shapiro}}]{zhuang2017fading}%
  \BibitemOpen
  \bibfield  {author} {\bibinfo {author} {\bibfnamefont {Q.}~\bibnamefont
  {Zhuang}}, \bibinfo {author} {\bibfnamefont {Z.}~\bibnamefont {Zhang}},\ and\
  \bibinfo {author} {\bibfnamefont {J.~H.}\ \bibnamefont {Shapiro}},\
  }\bibfield  {title} {\bibinfo {title} {Quantum illumination for enhanced
  detection of rayleigh-fading targets},\ }\href
  {https://doi.org/10.1103/PhysRevA.96.020302} {\bibfield  {journal} {\bibinfo
  {journal} {Phys. Rev. A}\ }\textbf {\bibinfo {volume} {96}},\ \bibinfo
  {pages} {020302} (\bibinfo {year} {2017}{\natexlab{c}})}\BibitemShut
  {NoStop}%
\bibitem [{\citenamefont {Barzanjeh}\ \emph {et~al.}(2015)\citenamefont
  {Barzanjeh}, \citenamefont {Guha}, \citenamefont {Weedbrook}, \citenamefont
  {Vitali}, \citenamefont {Shapiro},\ and\ \citenamefont
  {Pirandola}}]{barzanjeh2015microwave}%
  \BibitemOpen
  \bibfield  {author} {\bibinfo {author} {\bibfnamefont {S.}~\bibnamefont
  {Barzanjeh}}, \bibinfo {author} {\bibfnamefont {S.}~\bibnamefont {Guha}},
  \bibinfo {author} {\bibfnamefont {C.}~\bibnamefont {Weedbrook}}, \bibinfo
  {author} {\bibfnamefont {D.}~\bibnamefont {Vitali}}, \bibinfo {author}
  {\bibfnamefont {J.~H.}\ \bibnamefont {Shapiro}},\ and\ \bibinfo {author}
  {\bibfnamefont {S.}~\bibnamefont {Pirandola}},\ }\bibfield  {title} {\bibinfo
  {title} {Microwave quantum illumination},\ }\href@noop {} {\bibfield
  {journal} {\bibinfo  {journal} {Phys. Rev. Lett.}\ }\textbf {\bibinfo
  {volume} {114}},\ \bibinfo {pages} {080503} (\bibinfo {year}
  {2015})}\BibitemShut {NoStop}%
\bibitem [{\citenamefont {Barzanjeh}\ \emph {et~al.}(2020)\citenamefont
  {Barzanjeh}, \citenamefont {Pirandola}, \citenamefont {Vitali},\ and\
  \citenamefont {Fink}}]{barzanjeh2020microwave}%
  \BibitemOpen
  \bibfield  {author} {\bibinfo {author} {\bibfnamefont {S.}~\bibnamefont
  {Barzanjeh}}, \bibinfo {author} {\bibfnamefont {S.}~\bibnamefont
  {Pirandola}}, \bibinfo {author} {\bibfnamefont {D.}~\bibnamefont {Vitali}},\
  and\ \bibinfo {author} {\bibfnamefont {J.~M.}\ \bibnamefont {Fink}},\
  }\bibfield  {title} {\bibinfo {title} {Microwave quantum illumination using a
  digital receiver},\ }\href@noop {} {\bibfield  {journal} {\bibinfo  {journal}
  {Sci. Adv.}\ }\textbf {\bibinfo {volume} {6}},\ \bibinfo {pages} {eabb0451}
  (\bibinfo {year} {2020})}\BibitemShut {NoStop}%
\bibitem [{\citenamefont {Chang}\ \emph {et~al.}(2019)\citenamefont {Chang},
  \citenamefont {Vadiraj}, \citenamefont {Bourassa}, \citenamefont {Balaji},\
  and\ \citenamefont {Wilson}}]{chang2019quantum}%
  \BibitemOpen
  \bibfield  {author} {\bibinfo {author} {\bibfnamefont {C.~S.}\ \bibnamefont
  {Chang}}, \bibinfo {author} {\bibfnamefont {A.}~\bibnamefont {Vadiraj}},
  \bibinfo {author} {\bibfnamefont {J.}~\bibnamefont {Bourassa}}, \bibinfo
  {author} {\bibfnamefont {B.}~\bibnamefont {Balaji}},\ and\ \bibinfo {author}
  {\bibfnamefont {C.}~\bibnamefont {Wilson}},\ }\bibfield  {title} {\bibinfo
  {title} {Quantum-enhanced noise radar},\ }\href@noop {} {\bibfield  {journal}
  {\bibinfo  {journal} {Appl. Phys. Lett.}\ }\textbf {\bibinfo {volume}
  {114}},\ \bibinfo {pages} {112601} (\bibinfo {year} {2019})}\BibitemShut
  {NoStop}%
\bibitem [{\citenamefont {Shapiro}(2020)}]{shapiro2020quantum}%
  \BibitemOpen
  \bibfield  {author} {\bibinfo {author} {\bibfnamefont {J.~H.}\ \bibnamefont
  {Shapiro}},\ }\bibfield  {title} {\bibinfo {title} {The quantum illumination
  story},\ }\href@noop {} {\bibfield  {journal} {\bibinfo  {journal} {IEEE
  Trans. Aerosp. Electron. Syst.}\ }\textbf {\bibinfo {volume} {35}},\ \bibinfo
  {pages} {8} (\bibinfo {year} {2020})}\BibitemShut {NoStop}%
\bibitem [{\citenamefont {Zhuang}\ and\ \citenamefont
  {Pirandola}(2020{\natexlab{a}})}]{zhuang2020entanglement}%
  \BibitemOpen
  \bibfield  {author} {\bibinfo {author} {\bibfnamefont {Q.}~\bibnamefont
  {Zhuang}}\ and\ \bibinfo {author} {\bibfnamefont {S.}~\bibnamefont
  {Pirandola}},\ }\bibfield  {title} {\bibinfo {title} {Entanglement-enhanced
  testing of multiple quantum hypotheses},\ }\href@noop {} {\bibfield
  {journal} {\bibinfo  {journal} {Commun. Phys.}\ }\textbf {\bibinfo {volume}
  {3}},\ \bibinfo {pages} {1} (\bibinfo {year}
  {2020}{\natexlab{a}})}\BibitemShut {NoStop}%
\bibitem [{\citenamefont {Zhuang}\ and\ \citenamefont
  {Pirandola}(2020{\natexlab{b}})}]{zhuang2020ultimate}%
  \BibitemOpen
  \bibfield  {author} {\bibinfo {author} {\bibfnamefont {Q.}~\bibnamefont
  {Zhuang}}\ and\ \bibinfo {author} {\bibfnamefont {S.}~\bibnamefont
  {Pirandola}},\ }\bibfield  {title} {\bibinfo {title} {Ultimate limits for
  multiple quantum channel discrimination},\ }\href
  {https://doi.org/10.1103/PhysRevLett.125.080505} {\bibfield  {journal}
  {\bibinfo  {journal} {Phys. Rev. Lett.}\ }\textbf {\bibinfo {volume} {125}},\
  \bibinfo {pages} {080505} (\bibinfo {year} {2020}{\natexlab{b}})}\BibitemShut
  {NoStop}%
\bibitem [{\citenamefont {Karsa}\ and\ \citenamefont
  {Pirandola}(2020)}]{karsa2020energetic}%
  \BibitemOpen
  \bibfield  {author} {\bibinfo {author} {\bibfnamefont {A.}~\bibnamefont
  {Karsa}}\ and\ \bibinfo {author} {\bibfnamefont {S.}~\bibnamefont
  {Pirandola}},\ }\bibfield  {title} {\bibinfo {title} {Energetic
  considerations in quantum target ranging},\ }\href@noop {} {\bibfield
  {journal} {\bibinfo  {journal} {arXiv:2011.03637}\ } (\bibinfo {year}
  {2020})}\BibitemShut {NoStop}%
\bibitem [{\citenamefont {Weedbrook}\ \emph {et~al.}(2012)\citenamefont
  {Weedbrook}, \citenamefont {Pirandola}, \citenamefont {Garc\'{\i}a-Patr\'on},
  \citenamefont {Cerf}, \citenamefont {Ralph}, \citenamefont {Shapiro},\ and\
  \citenamefont {Lloyd}}]{Weedbrook2012}%
  \BibitemOpen
  \bibfield  {author} {\bibinfo {author} {\bibfnamefont {C.}~\bibnamefont
  {Weedbrook}}, \bibinfo {author} {\bibfnamefont {S.}~\bibnamefont
  {Pirandola}}, \bibinfo {author} {\bibfnamefont {R.}~\bibnamefont
  {Garc\'{\i}a-Patr\'on}}, \bibinfo {author} {\bibfnamefont {N.~J.}\
  \bibnamefont {Cerf}}, \bibinfo {author} {\bibfnamefont {T.~C.}\ \bibnamefont
  {Ralph}}, \bibinfo {author} {\bibfnamefont {J.~H.}\ \bibnamefont {Shapiro}},\
  and\ \bibinfo {author} {\bibfnamefont {S.}~\bibnamefont {Lloyd}},\ }\bibfield
   {title} {\bibinfo {title} {Gaussian quantum information},\ }\href
  {https://doi.org/10.1103/RevModPhys.84.621} {\bibfield  {journal} {\bibinfo
  {journal} {Rev. Mod. Phys.}\ }\textbf {\bibinfo {volume} {84}},\ \bibinfo
  {pages} {621} (\bibinfo {year} {2012})}\BibitemShut {NoStop}%
\bibitem [{\citenamefont {Pirandola}(2011)}]{pirandola2011quantum}%
  \BibitemOpen
  \bibfield  {author} {\bibinfo {author} {\bibfnamefont {S.}~\bibnamefont
  {Pirandola}},\ }\bibfield  {title} {\bibinfo {title} {Quantum reading of a
  classical digital memory},\ }\href@noop {} {\bibfield  {journal} {\bibinfo
  {journal} {Phys. Rev. Lett.}\ }\textbf {\bibinfo {volume} {106}},\ \bibinfo
  {pages} {090504} (\bibinfo {year} {2011})}\BibitemShut {NoStop}%
\bibitem [{\citenamefont {Nair}\ and\ \citenamefont
  {Gu}(2020)}]{nair2020fundamental}%
  \BibitemOpen
  \bibfield  {author} {\bibinfo {author} {\bibfnamefont {R.}~\bibnamefont
  {Nair}}\ and\ \bibinfo {author} {\bibfnamefont {M.}~\bibnamefont {Gu}},\
  }\bibfield  {title} {\bibinfo {title} {Fundamental limits of quantum
  illumination},\ }\href@noop {} {\bibfield  {journal} {\bibinfo  {journal}
  {Optica}\ }\textbf {\bibinfo {volume} {7}},\ \bibinfo {pages} {771} (\bibinfo
  {year} {2020})}\BibitemShut {NoStop}%
\bibitem [{\citenamefont {Bradshaw}\ \emph {et~al.}(2020)\citenamefont
  {Bradshaw}, \citenamefont {Conlon}, \citenamefont {Tserkis}, \citenamefont
  {Gu}, \citenamefont {Lam},\ and\ \citenamefont
  {Assad}}]{bradshaw2020optimal}%
  \BibitemOpen
  \bibfield  {author} {\bibinfo {author} {\bibfnamefont {M.}~\bibnamefont
  {Bradshaw}}, \bibinfo {author} {\bibfnamefont {L.~O.}\ \bibnamefont
  {Conlon}}, \bibinfo {author} {\bibfnamefont {S.}~\bibnamefont {Tserkis}},
  \bibinfo {author} {\bibfnamefont {M.}~\bibnamefont {Gu}}, \bibinfo {author}
  {\bibfnamefont {P.~K.}\ \bibnamefont {Lam}},\ and\ \bibinfo {author}
  {\bibfnamefont {S.~M.}\ \bibnamefont {Assad}},\ }\bibfield  {title} {\bibinfo
  {title} {Optimal probes for continuous variable quantum illumination},\
  }\href@noop {} {\bibfield  {journal} {\bibinfo  {journal} {arXiv:2010.09156}\
  } (\bibinfo {year} {2020})}\BibitemShut {NoStop}%
\bibitem [{\citenamefont {Li}(2016)}]{li2016discriminating}%
  \BibitemOpen
  \bibfield  {author} {\bibinfo {author} {\bibfnamefont {K.}~\bibnamefont
  {Li}},\ }\bibfield  {title} {\bibinfo {title} {Discriminating quantum states:
  The multiple chernoff distance},\ }\href@noop {} {\bibfield  {journal}
  {\bibinfo  {journal} {Ann. Statist.}\ }\textbf {\bibinfo {volume} {44}},\
  \bibinfo {pages} {1661} (\bibinfo {year} {2016})}\BibitemShut {NoStop}%
\bibitem [{\citenamefont {Nussbaum}\ \emph {et~al.}(2011)\citenamefont
  {Nussbaum}, \citenamefont {Szko{\l}a} \emph
  {et~al.}}]{nussbaum2011asymptotic}%
  \BibitemOpen
  \bibfield  {author} {\bibinfo {author} {\bibfnamefont {M.}~\bibnamefont
  {Nussbaum}}, \bibinfo {author} {\bibfnamefont {A.}~\bibnamefont {Szko{\l}a}},
  \emph {et~al.},\ }\bibfield  {title} {\bibinfo {title} {An asymptotic error
  bound for testing multiple quantum hypotheses},\ }\href@noop {} {\bibfield
  {journal} {\bibinfo  {journal} {Ann. Statist.}\ }\textbf {\bibinfo {volume}
  {39}},\ \bibinfo {pages} {3211} (\bibinfo {year} {2011})}\BibitemShut
  {NoStop}%
\bibitem [{\citenamefont {Audenaert}\ \emph {et~al.}(2007)\citenamefont
  {Audenaert}, \citenamefont {Calsamiglia}, \citenamefont {Munoz-Tapia},
  \citenamefont {Bagan}, \citenamefont {Masanes}, \citenamefont {Acin},\ and\
  \citenamefont {Verstraete}}]{audenaert2007discriminating}%
  \BibitemOpen
  \bibfield  {author} {\bibinfo {author} {\bibfnamefont {K.~M.}\ \bibnamefont
  {Audenaert}}, \bibinfo {author} {\bibfnamefont {J.}~\bibnamefont
  {Calsamiglia}}, \bibinfo {author} {\bibfnamefont {R.}~\bibnamefont
  {Munoz-Tapia}}, \bibinfo {author} {\bibfnamefont {E.}~\bibnamefont {Bagan}},
  \bibinfo {author} {\bibfnamefont {L.}~\bibnamefont {Masanes}}, \bibinfo
  {author} {\bibfnamefont {A.}~\bibnamefont {Acin}},\ and\ \bibinfo {author}
  {\bibfnamefont {F.}~\bibnamefont {Verstraete}},\ }\bibfield  {title}
  {\bibinfo {title} {Discriminating states: The quantum chernoff bound},\
  }\href@noop {} {\bibfield  {journal} {\bibinfo  {journal} {Phys. Rev. Lett.}\
  }\textbf {\bibinfo {volume} {98}},\ \bibinfo {pages} {160501} (\bibinfo
  {year} {2007})}\BibitemShut {NoStop}%
\bibitem [{\citenamefont {Pirandola}\ and\ \citenamefont
  {Lloyd}(2008)}]{Pirandola2008}%
  \BibitemOpen
  \bibfield  {author} {\bibinfo {author} {\bibfnamefont {S.}~\bibnamefont
  {Pirandola}}\ and\ \bibinfo {author} {\bibfnamefont {S.}~\bibnamefont
  {Lloyd}},\ }\bibfield  {title} {\bibinfo {title} {Computable bounds for the
  discrimination of gaussian states},\ }\href
  {https://doi.org/10.1103/PhysRevA.78.012331} {\bibfield  {journal} {\bibinfo
  {journal} {Phys. Rev. A}\ }\textbf {\bibinfo {volume} {78}},\ \bibinfo
  {pages} {012331} (\bibinfo {year} {2008})}\BibitemShut {NoStop}%
\bibitem [{\citenamefont {Helstrom}(1976)}]{Helstrom_1976}%
  \BibitemOpen
  \bibfield  {author} {\bibinfo {author} {\bibfnamefont {C.}~\bibnamefont
  {Helstrom}},\ }\href {https://books.google.com/books?id=fv9SAAAAMAAJ} {\emph
  {\bibinfo {title} {Quantum Detection and Estimation Theory}}},\ Mathematics
  in Science and Engineering : a series of monographs and textbooks\ (\bibinfo
  {publisher} {Academic Press},\ \bibinfo {year} {1976})\BibitemShut {NoStop}%
\bibitem [{\citenamefont {Cariolaro}\ and\ \citenamefont
  {Pierobon}(2010)}]{cariolaro2010theory}%
  \BibitemOpen
  \bibfield  {author} {\bibinfo {author} {\bibfnamefont {G.}~\bibnamefont
  {Cariolaro}}\ and\ \bibinfo {author} {\bibfnamefont {G.}~\bibnamefont
  {Pierobon}},\ }\bibfield  {title} {\bibinfo {title} {Theory of quantum pulse
  position modulation and related numerical problems},\ }\href@noop {}
  {\bibfield  {journal} {\bibinfo  {journal} {IEEE Trans. Commun.}\ }\textbf
  {\bibinfo {volume} {58}},\ \bibinfo {pages} {1213} (\bibinfo {year}
  {2010})}\BibitemShut {NoStop}%
\bibitem [{\citenamefont {Holevo}(1978)}]{PGM1}%
  \BibitemOpen
  \bibfield  {author} {\bibinfo {author} {\bibfnamefont {A.~S.}\ \bibnamefont
  {Holevo}},\ }\bibfield  {title} {\bibinfo {title} {On asymptotically optimal
  hypotheses testing in quantum statistics},\ }\href@noop {} {\bibfield
  {journal} {\bibinfo  {journal} {Teoriya Veroyatnostei i ee Primeneniya}\
  }\textbf {\bibinfo {volume} {23}},\ \bibinfo {pages} {429} (\bibinfo {year}
  {1978})}\BibitemShut {NoStop}%
\bibitem [{\citenamefont {Hausladen}\ and\ \citenamefont
  {Wootters}(1994)}]{PGM2}%
  \BibitemOpen
  \bibfield  {author} {\bibinfo {author} {\bibfnamefont {P.}~\bibnamefont
  {Hausladen}}\ and\ \bibinfo {author} {\bibfnamefont {W.~K.}\ \bibnamefont
  {Wootters}},\ }\bibfield  {title} {\bibinfo {title} {A ‘pretty
  good’measurement for distinguishing quantum states},\ }\href@noop {}
  {\bibfield  {journal} {\bibinfo  {journal} {J. Mod. Opt.}\ }\textbf {\bibinfo
  {volume} {41}},\ \bibinfo {pages} {2385} (\bibinfo {year}
  {1994})}\BibitemShut {NoStop}%
\bibitem [{\citenamefont {Hausladen}\ \emph
  {et~al.}(1996{\natexlab{a}})\citenamefont {Hausladen}, \citenamefont {Jozsa},
  \citenamefont {Schumacher}, \citenamefont {Westmoreland},\ and\ \citenamefont
  {Wootters}}]{PGM3}%
  \BibitemOpen
  \bibfield  {author} {\bibinfo {author} {\bibfnamefont {P.}~\bibnamefont
  {Hausladen}}, \bibinfo {author} {\bibfnamefont {R.}~\bibnamefont {Jozsa}},
  \bibinfo {author} {\bibfnamefont {B.}~\bibnamefont {Schumacher}}, \bibinfo
  {author} {\bibfnamefont {M.}~\bibnamefont {Westmoreland}},\ and\ \bibinfo
  {author} {\bibfnamefont {W.~K.}\ \bibnamefont {Wootters}},\ }\bibfield
  {title} {\bibinfo {title} {Classical information capacity of a quantum
  channel},\ }\href {https://doi.org/10.1103/PhysRevA.54.1869} {\bibfield
  {journal} {\bibinfo  {journal} {Phys. Rev. A}\ }\textbf {\bibinfo {volume}
  {54}},\ \bibinfo {pages} {1869} (\bibinfo {year}
  {1996}{\natexlab{a}})}\BibitemShut {NoStop}%
\bibitem [{\citenamefont {Banchi}\ \emph {et~al.}(2015)\citenamefont {Banchi},
  \citenamefont {Braunstein},\ and\ \citenamefont {Pirandola}}]{banchi2015}%
  \BibitemOpen
  \bibfield  {author} {\bibinfo {author} {\bibfnamefont {L.}~\bibnamefont
  {Banchi}}, \bibinfo {author} {\bibfnamefont {S.~L.}\ \bibnamefont
  {Braunstein}},\ and\ \bibinfo {author} {\bibfnamefont {S.}~\bibnamefont
  {Pirandola}},\ }\bibfield  {title} {\bibinfo {title} {Quantum fidelity for
  arbitrary gaussian states},\ }\href
  {https://doi.org/10.1103/PhysRevLett.115.260501} {\bibfield  {journal}
  {\bibinfo  {journal} {Phys. Rev. Lett.}\ }\textbf {\bibinfo {volume} {115}},\
  \bibinfo {pages} {260501} (\bibinfo {year} {2015})}\BibitemShut {NoStop}%
\bibitem [{\citenamefont {Hausladen}\ \emph
  {et~al.}(1996{\natexlab{b}})\citenamefont {Hausladen}, \citenamefont {Jozsa},
  \citenamefont {Schumacher}, \citenamefont {Westmoreland},\ and\ \citenamefont
  {Wootters}}]{hausladen1996classical}%
  \BibitemOpen
  \bibfield  {author} {\bibinfo {author} {\bibfnamefont {P.}~\bibnamefont
  {Hausladen}}, \bibinfo {author} {\bibfnamefont {R.}~\bibnamefont {Jozsa}},
  \bibinfo {author} {\bibfnamefont {B.}~\bibnamefont {Schumacher}}, \bibinfo
  {author} {\bibfnamefont {M.}~\bibnamefont {Westmoreland}},\ and\ \bibinfo
  {author} {\bibfnamefont {W.~K.}\ \bibnamefont {Wootters}},\ }\bibfield
  {title} {\bibinfo {title} {Classical information capacity of a quantum
  channel},\ }\href@noop {} {\bibfield  {journal} {\bibinfo  {journal} {Phys.
  Rev. A}\ }\textbf {\bibinfo {volume} {54}},\ \bibinfo {pages} {1869}
  (\bibinfo {year} {1996}{\natexlab{b}})}\BibitemShut {NoStop}%
\bibitem [{\citenamefont {Schumacher}\ and\ \citenamefont
  {Westmoreland}(1997)}]{schumacher1997sending}%
  \BibitemOpen
  \bibfield  {author} {\bibinfo {author} {\bibfnamefont {B.}~\bibnamefont
  {Schumacher}}\ and\ \bibinfo {author} {\bibfnamefont {M.~D.}\ \bibnamefont
  {Westmoreland}},\ }\bibfield  {title} {\bibinfo {title} {Sending classical
  information via noisy quantum channels},\ }\href@noop {} {\bibfield
  {journal} {\bibinfo  {journal} {Phys. Rev. A}\ }\textbf {\bibinfo {volume}
  {56}},\ \bibinfo {pages} {131} (\bibinfo {year} {1997})}\BibitemShut
  {NoStop}%
\bibitem [{\citenamefont {Holevo}(1998)}]{holevo1998capacity}%
  \BibitemOpen
  \bibfield  {author} {\bibinfo {author} {\bibfnamefont {A.~S.}\ \bibnamefont
  {Holevo}},\ }\bibfield  {title} {\bibinfo {title} {The capacity of the
  quantum channel with general signal states},\ }\href@noop {} {\bibfield
  {journal} {\bibinfo  {journal} {IEEE Trans. Inf. Theory}\ }\textbf {\bibinfo
  {volume} {44}},\ \bibinfo {pages} {269} (\bibinfo {year} {1998})}\BibitemShut
  {NoStop}%
\bibitem [{\citenamefont {Giovannetti}\ \emph {et~al.}(2014)\citenamefont
  {Giovannetti}, \citenamefont {Garcia-Patron}, \citenamefont {Cerf},\ and\
  \citenamefont {Holevo}}]{giovannetti2014ultimate}%
  \BibitemOpen
  \bibfield  {author} {\bibinfo {author} {\bibfnamefont {V.}~\bibnamefont
  {Giovannetti}}, \bibinfo {author} {\bibfnamefont {R.}~\bibnamefont
  {Garcia-Patron}}, \bibinfo {author} {\bibfnamefont {N.~J.}\ \bibnamefont
  {Cerf}},\ and\ \bibinfo {author} {\bibfnamefont {A.~S.}\ \bibnamefont
  {Holevo}},\ }\bibfield  {title} {\bibinfo {title} {Ultimate classical
  communication rates of quantum optical channels},\ }\href@noop {} {\bibfield
  {journal} {\bibinfo  {journal} {Nature Photonics}\ }\textbf {\bibinfo
  {volume} {8}},\ \bibinfo {pages} {796} (\bibinfo {year} {2014})}\BibitemShut
  {NoStop}%
\bibitem [{\citenamefont {Zhuang}\ \emph
  {et~al.}(2017{\natexlab{d}})\citenamefont {Zhuang}, \citenamefont {Zhang},\
  and\ \citenamefont {Shapiro}}]{zhuang2017entanglement}%
  \BibitemOpen
  \bibfield  {author} {\bibinfo {author} {\bibfnamefont {Q.}~\bibnamefont
  {Zhuang}}, \bibinfo {author} {\bibfnamefont {Z.}~\bibnamefont {Zhang}},\ and\
  \bibinfo {author} {\bibfnamefont {J.~H.}\ \bibnamefont {Shapiro}},\
  }\bibfield  {title} {\bibinfo {title} {Entanglement-enhanced lidars for
  simultaneous range and velocity measurements},\ }\href@noop {} {\bibfield
  {journal} {\bibinfo  {journal} {Phys. Rev. A}\ }\textbf {\bibinfo {volume}
  {96}},\ \bibinfo {pages} {040304} (\bibinfo {year}
  {2017}{\natexlab{d}})}\BibitemShut {NoStop}%
\end{thebibliography}
%apsrev4-2.bst 2019-01-14 (MD) hand-edited version of apsrev4-1.bst
%Control: key (0)
%Control: author (8) initials jnrlst
%Control: editor formatted (1) identically to author
%Control: production of article title (0) allowed
%Control: page (0) single
%Control: year (1) truncated
%Control: production of eprint (0) enabled
%

\end{document}